\documentclass[twocolumn,showpacs,preprintnumbers,amsmath,amssymb]{revtex4}

\usepackage{graphicx}
\usepackage{xcolor}
\definecolor{indigo(dye)}{rgb}{0.0,0.25,0.42}

\begin{document}

\title{Capped carbon nanotubes with a number of ground state magnetization discontinuities increasing with their size\footnote{Dedicated to the memory of Dr. Brahim Belhadji.}}

\author{N. P. Konstantinidis}
\affiliation{Leoforos Syggrou 360, Kallithea 17674, Athens, Hellas}

\date{\today}

\begin{abstract}
The classical ground state magnetic response of fullerene molecules that resemble capped carbon nanotubes is calculated within the framework of the antiferromagnetic Heisenberg model. It is found that the magnetic response depends subtly on spatial symmetry and chirality.
Clusters based on armchair carbon nanotubes which are capped with non-neighboring pentagons and have $D_{5d}$ spatial symmetry have a number of magnetization discontinuities which increases with their size. This occurs even though the model completely lacks magnetic anisotropy, and even though the only source of frustration are the two groups of six pentagons located at the ends of the molecules, which become more strongly outnumbered as the clusters are filled in the middle with more unfrustrated hexagons with increasing size. For the cluster with 180 vertices there are already seven magnetization and one susceptibility discontinuities. Contrary to that, similar molecules which have $D_{5h}$ spatial symmetry reach a limit of one magnetization and two susceptibility ground state discontinuities, while fullerene molecules based on zigzag carbon nanotubes and capped by neighboring pentagons also reach a fixed number of discontinuities with increasing size.
\end{abstract}


\maketitle

\section{Introduction}
\label{sec:introduction}

Fullerenes are carbon structures that come among others in the form of hollow spheres (molecules) or nanotubes \cite{Fowler95}. Fullerene molecules consist of 12 pentagons and $\frac{N}{2}-10$ hexagons which share edges, with $N$ the number of the molecule's vertices, which are three-fold coordinated. A pentagon ring that hosts a spin on each vertex and has nearest-neighbor spins interacting antiferromagnetically is frustrated when isolated, unlike a hexagon. This is because classical nearest-neighbor spins are not antiparallel in the lowest energy configuration of the pentagon, in contrast to what happens for a hexagon. As a result, pentagons introduce frustration in a fullerene molecule and act as structural defects. Since the pentagon number is fixed frustration is stronger for smaller $N$, where the pentagons outnumber or are comparable in number to the hexagons and also neighbor each other. The influence of frustration is however significant for any $N$. As $N$ goes to infinity the fullerene molecules resemble more and more a two dimensional hexagonal lattice with periodic boundary conditions, whose properties depend on the location of the 12 pentagonal structural defects. Lattices and clusters of frustrated topology are of special interest, as
they are associated with unexpected magnetic behavior \cite{Anderson73,Diep05,Schnack10,NPK01,NPK02,NPK04,NPK13}.

Perhaps the most well-known fullerene molecule is $C_{60}$. It has the shape of the truncated icosahedron and belongs to the icosahedral symmetry group $I_h$ \cite{Altmann94}. $C_{60}$ superconducts when doped with alkali metals \cite{Hebard91}. For the undoped $C_{60}$ solid, band theory predicts that the $h_u$ band is full while the $t_{1u}$ band is empty, thus it is an insulator \cite{Gunnarsson97}. When solid $C_{60}$ is doped by alkali atoms $A$, each atom donates about one electron to the $t_{1u}$ band \cite{Saito91,Erwin91}. The $t_{1u}$ band can accommodate six electrons, however $A_4C_{60}$ turns out to be an insulator \cite{Kiefl92,Murphy92,Benning93}, which shows the importance of electron-electron correlations. Strong electron correlations have been treated with LDA band structure calculations \cite{Yang03}, but it has been desirable to consider electron correlations within the more accurate framework of the Hubbard model \cite{Chakravarty91,Lin07}.

A first approximation in a fullerene molecule has three of each carbon atom's valence electrons forming three $\sigma$-bonds using $sp^2$ hybrid orbitals, while the fourth valence electron is delocalized.
Thus each carbon has one active radial p-orbital. When the long-range Coulomb or the on-site Hubbard repulsion $U$ is much stronger than the nearest neighbor Hubbard hopping $t$ \cite{Auerbach98,Fazekas99}, there is essentially one electron in each orbital and the interaction between nearest neighbors is antiferromagnetic \cite{Coffey92}. The antiferromagnetic Heisenberg model (AHM) becomes of relevance then.
Estimates for the on-site repulsion $U$ show that $C_{60}$ belongs to the intermediate $U$ regime of the Hubbard model \cite{Chakravarty91,Stollhoff91}.

In an external field the classical magnetization usually increases continuously with the field, and the susceptibility is a smooth function of it. It has however been found that the spin-flop phenomenon can lead to a discontinuous magnetization response \cite{Neel36}. This phenomenon is associated with spin space anisotropy, with the energy more efficiently minimized along certain directions in spin space. It is of particular interest when specific geometries support jumps in the magnetization and the susceptibility when there is no anisotropy in the magnetic interactions. This shows for example the possibility to tune the magnetization between well-separated values with fine changes of an applied field. In such cases the origin of the discontinuity lies in the frustrated connectivity of the magnetic interactions, which reflects the topology of the structure that hosts the magnetic units.

Within the framework of the AHM, which lacks any magnetic anisotropy, the magnetization response of the $I_h$ symmetry fullerenes is discontinuous at the classical and quantum level \cite{Coffey92,NPK05,NPK07}.
For relatively small fullerene molecules of other symmetries the classical ground state magnetization curve is also rich with discontinuities, while there are only pronounced magnetization plateaus for $s=\frac{1}{2}$ \cite{NPK17,NPK09}. In addition the icosahedron, not a fullerene but the smallest cluster with $I_h$ symmetry, has a classical ground state magnetization discontinuity which persists for lower values of $s$ \cite{Schroeder05,NPK15}.
The addition of the next order isotropic term in the Hamiltonian, the biquadratic exchange interaction, significantly enriches the icosahedron's magnetization response with multiple magnetization and susceptibility discontinuities \cite{NPK16-1}. The above results show strong correlations between spatial symmetry and magnetic response for a fullerene molecule, which have also been demonstrated for the dodecahedron and the icosahedron \cite{NPK05}. It it also noted that in numerical calculations $U$ has been found to be stronger for C$_{20}$ than C$_{60}$ \cite{Lin07,Lin08}, further supporting the validity of the AHM as a good description of the Hubbard model for the $I_h$ fullerenes.

Carbon nanotubes are cylindrically structured allotropes of carbon that form aesthetically pleasing hexagonal shapes, with each carbon atom covalently bonded to other three \cite{Iijima91,Saito98,Dekker99,Terrones03,DeVolder13}. Their topology is the one of the fullerene family, and they can be viewed as rolled-up sheets of graphene. They possess a large number of unusual mechanical, electrical and thermal properties \cite{Peng08,Pop06}. Their diameter can be quite smaller than a nanometer \cite{Hayashi03} and their length as big as several centimeters, bridging the molecular and macroscopic scales. The ends of the nanotubes can be closed by carbon atoms that form pentagons at the ends of the nanotube \cite{deJonge05}.
Their conducting properties are strongly dependent on the chirality and diameter of the hexagonal lattice along the nanotube. Armchair carbon nanotubes are metallic and can support very high currents, while zigzag and chiral carbon nanotubes are semiconducting \cite{Wei01,Mintmire92,Hamada92,Saito92}.

Single-wall carbon nanotubes are ideal one-dimensional mesoscopic systems and can behave as quantum waveguides \cite{Liang01} and quantum dots \cite{Tans97,Bockrath97,Cobden02}. Oxygen molecules encapsulated in single-wall carbon nanotubes form a one-dimensional $s=1$ Haldane magnet \cite{Hagiwara14}. Vacancies and atom substitutions in nanotubes have been shown to lead to magnetic coupling \cite{Singh13,Zhang11}. Theoretical calculations showed that ferromagnetism can be induced by carbon adatoms on boron nitride nanotubes \cite{Zhou09}. Fluorine atoms topologically adsorbed on nanotubes of boron nitride induce long-ranged ferromagnetic ordering along the tube \cite{Zhang09}. Hydrogen trapped at a zigzag nanotube vacancy and hydrogen bonded to a carbon adatom on the surface of a nanotube of any diameter and chirality lead to the spin polarization of delocalized $\pi$ electrons \cite{Ma05}. Magnetic impurities which are doped in zigzag carbon nanotubes change the semiconducting state into a half metallic magnetic state \cite{Moradian06}. Carbon nanotubes become magnetized when in contact with a magnetic material. Spin-polarized charge transfer at the interface between a flat ferromagnetic metal substrate and a multiwalled nanotube leads to a spin transfer of about 0.1 $\mu_B$ per contact carbon atom \cite{Cespedes04}.

From the theoretical point of view, the electronic properties of the nanotubes are closely connected to the ones of graphene \cite{Geim07,Neto09}. Ideal graphene is nonmagnetic, but many of its derivatives are. The nearest-neighbor tight-binding model has been shown to describe accurately the electronic structure of graphene and derivatives of it \cite{Yazyev10}. In the absence of electronic interactions graphene has a linear dispersion in the vicinity of the Dirac points.
However experiments have highlighted the importance of electron interactions which induce gaps at the Dirac points \cite{Deshpande09,Bostwick07}. The low-dimensionality of the nanotubes also qualifies them as a good testing ground for strongly correlated interactions \cite{Kane97,Bockrath99}. Due to experimental difficulties electron correlations effects have been difficult to observe, and a tight-binding picture has also been adopted. However experiments with very clean samples highlighted the importance of electron correlations \cite{Siegel13,Freitag12,Zhou08,Mayorov11,Luu16}, which have been taken into account in the Hubbard model with on-site and nearest-neighbor interactions \cite{Egger97,Lin98,Yoshioka99,Nersesyan03,Balents97,Krotov97,Lin98-1,Bunder08,Lopez01,Zhao04,Alfonsi10}. The intermediate $U$ regime of the Hubbard model has emerged as the most important for the nanotubes.

Motivated by the subtle dependence of the conductive properties of carbon nanotubes on their topology and by the magnetization discontinuities of the aforementioned fullerene molecules, in this paper we undertake the question of the magnetic response of fullerene molecules that resemble capped carbon nanotubes. They have their cylindrical structure and are capped at their ends by clusters that contain six pentagons, introducing frustration. The first family has the structure of armchair nanotubes, with a cap that is one half of the truncated icosahedron $C_{60}$, where pentagons do not neighbor each other and the central pentagon is located at the apex (Fig. \ref{fig:isolatedpentagonssubcluster}(a)) \cite{Yang10}. These structures provide a link between molecular fullerenes and carbon nanotubes, being capped versions of (5,5) single-walled armchair carbon nanotubes. The smallest member of the class is C$_{60}$, while bigger molecules have either a number of vertices which is an even multiple of 10 and $D_{5d}$ spatial symmetry (Fig. \ref{fig:armchair}(a)), or a number of vertices which is an odd multiple of 10 and $D_{5h}$ spatial symmetry (Fig. \ref{fig:armchair}(b)). As $N$ increases so does the length of the body of the molecule with its nanotube shape \cite{Fowler95}, but not its diameter. The basic question relates to the influence of the pentagon-including caps at the edges on the magnetic properties and if it diminishes with $N$. It has already been shown that an increasing size does not alter the magnetic properties of the $I_h$ molecules originating in the 12 frustrated pentagons \cite{NPK07}. In this paper it is demonstrated that for the $D_{5d}$ symmetry members of this particular class of fullerene molecules resembling (5,5) armchair carbon nanotubes the magnetic response becomes even richer with $N$.


We consider classical spins mounted on the vertices of the molecules interacting according to the AHM in order to examine correlations between spatial symmetry and the number of ground state magnetization and susceptibility discontinuities in a field.
We find that the $D_{5d}$ molecules based on (5,5) armchair nanotubes have a number of ground state discontinuities which increases with their size, and goes up to seven magnetization and one susceptibility jumps for the largest cluster considered, which has 180 vertices. This number far supersedes the three magnetization gaps of the dodecahedron, and the results indicate that the total number of ground state discontinuities is controlled by the size of the molecule and will keep increasing for bigger members of this class. On the other hand, for the $D_{5h}$ symmetry clusters based on (5,5) armchair nanotubes the number of discontinuities is equal to three for all the calculated sizes, with one magnetization and two susceptibility discontinuities for the largest sizes. This result shows that the spatial symmetry of the molecule plays a very important role in the magnetic response.

It is also demonstrated by considering a second class of fullerene molecules that apart from spatial symmetry the chirality of the nanotube is very important for its magnetic response. These molecules have a single-wall (5,0) zigzag carbon nanotube shape and are capped at their ends with neighboring pentagons, which are the part of the dodecahedron that remains after one of its pentagons is removed (Fig. \ref{fig:nonisolatedpentagonssubcluster}(a)). They also alternate their symmetry between $D_{5d}$ and $D_{5h}$ whenever ten new vertices are added to the molecule (Fig. \ref{fig:zigzag}). In contrast to the (5,5) armchair molecules they have a size independent total number of discontinuities irrespectively of their spatial symmetry. This shows that fine differences in chirality and topology can have a significant effect on the magnetic properties of the nanotube-like fullerene molecules.

In the case of the icosahedron it was shown that the classical magnetization discontinuity originates in the particular structure of the cluster, being made up of a closed strip of a triangular lattice plus two extra vertices whose spins respond immediately to the field when not linked with the rest of the cluster \cite{NPK15}. A similar procedure for the molecules in this paper is to calculate the magnetic response of the caps and the nanotube body separately, and then introduce their coupling until all exchange constants of the fullerene molecule become equal.
This is a numerically tedious procedure and was done for the $N=80$ molecule of (5,5) armchair-type, where in general the discontinuities can not be traced back directly to the ones of the caps in a one-to-one fashion. At the same time it is shown that the magnetic response of the (5,5) armchair and (5,0) zigzag caps is different, indicating the importance of the cap in the magnetization curve. It is concluded that the magnetic response depends on a fine competition of the chirality of the body of the molecule, the molecule's spatial symmetry, and also the nature of the cap.

The plan of this paper is as follows: in Sec. \ref{sec:model} the model is presented.
Section \ref{sec:armchairnanotubescappednonneighboringpentagons} deals with the magnetic response of (5,5) armchair nanotubes capped with non-neighboring pentagons,
and of (5,0) zigzag nanotubes capped with neighboring pentagons. Finally
Sec. \ref{sec:conclusions} presents the conclusions.

\section{Model}
\label{sec:model}

The Hamiltonian of the AHM is
\begin{equation}
H = J \sum_{<ij>} \textrm{} \vec{s}_i \cdot \vec{s}_j - h \sum_{i=1}^N s_i^z
\label{eqn:Hamiltonian}
\end{equation}
$J$ is the exchange interaction strength,
taken to be 1 from now on.
$<ij>$ indicates that interactions are limited to nearest neighbors $i$ and $j$. The magnetic field $\vec{h}$ is taken to be directed along the $\hat{z}$ axis, and the spins are classical vectors with unit magnitude. In Hamiltonian (\ref{eqn:Hamiltonian}) the exchange energy competes with the magnetic energy, with the frustrated connectivity playing an important role. Minimization of the Hamiltonian gives the ground state energy and spin configuration for any magnetic field \cite{Coffey92,NPK05,NPK07,NPK15-1,NPK15,NPK16,NPK17}.

Here we consider the case where the exchange interaction is the same for any spin pair, whether it belongs in the nanotube part or the cap or it relates to a nanotube spin and one in the cap. We do so in order to highlight the importance of the fullerene molecule connectivity for the magnetic properties. We also consider in Sec. \ref{subsec:D5dsymmetrymolecules} the (5,5) armchair-type molecule with $D_{5d}$ symmetry and $N=80$ in the case of varying couplings between the nanotube and the cap, in which case the magnetic diagram can become even richer, showing that the multitude of magnetization and susceptibility discontinuities is not the result of a special choice of exchange couplings. In addition, Sec. \ref{subsec:sixnonneighboringpentagons} demonstrates how the six non-neighboring pentagons subblock situated at the two edges of the (5,5) armchair-type molecules is a source of discontinuous magnetic response, irrespectively of its coupling strength with the nanotube part of the molecule. The same is illustrated in Sec. \ref{subsec:zigzagnanotubescappedneighboringpentagons} for the six neighboring pentagons subblock that caps the (5,0) zigzag-type molecules.


\section{(5,5) Armchair Nanotubes Capped with Non-Neighboring Pentagons}
\label{sec:armchairnanotubescappednonneighboringpentagons}

The geometry of a nanotube is determined by the chiral vector of the original hexagonal lattice, with nanotubes divided into armchair, zigzag and chiral. We consider (5,5) armchair nanotubes of varying size that are capped on each end by a sublock of six non-neighboring pentagons (Fig. \ref{fig:isolatedpentagonssubcluster}(a)), resulting in fullerene molecules. The spatial symmetry of these molecules depends on their number of vertices. When it is an even multiple of 10 the symmetry is $D_{5d}$ (Fig. \ref{fig:armchair}(a)), while for an odd multiple of 10 it changes to $D_{5h}$ (Fig. \ref{fig:armchair}(b)). Since in both cases the only source of magnetic frustration is the six pentagon subblock, its response in an external field when it is completely isolated is calculated.

\subsection{Six Non-Neighboring Pentagons}
\label{subsec:sixnonneighboringpentagons}

\begin{figure}
\centerline{
\includegraphics[width=1.6in,height=1.4in]{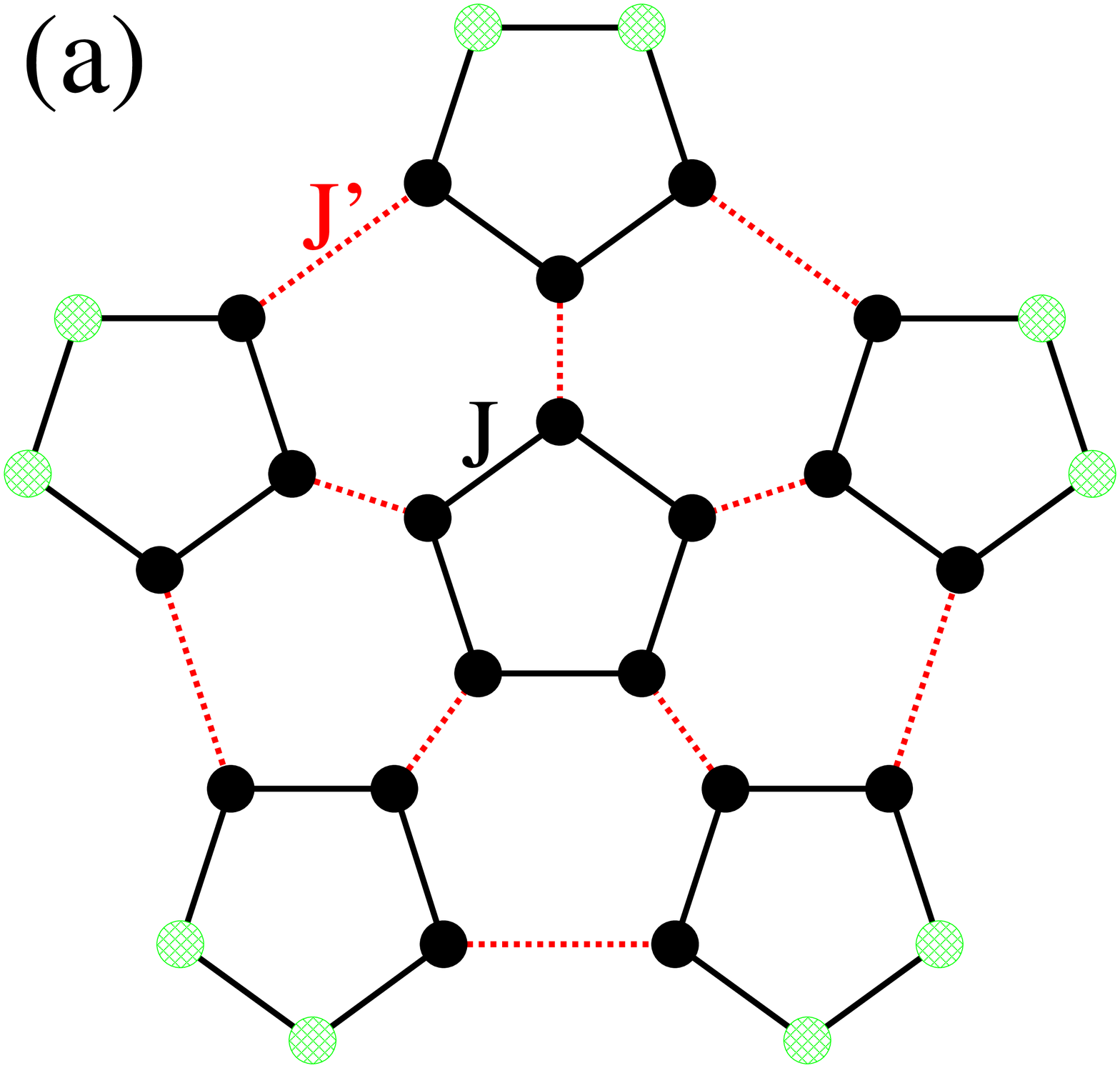}
\includegraphics[width=1.9in,height=1.4in]{Figure2}
}
\vspace{0pt}
\caption{(Color online) (a) Planar projection of the six isolated pentagons subblock that is the source of frustration for the $D_{5d}$ and $D_{5h}$ fullerene molecules resembling (5,5) armchair nanotubes. The circles are classical spins of unit magnitude, with the (green) patterned circles connecting the subblock with the nanotube part of the molecule. Each spin interacts with its nearest neighbors according to the connecting lines. The (black) solid and (red) dashed lines indicate intrapentagon interactions of strength $J$ and interpentagon interactions of strength $J'$ respectively. $J'$ varies from 0 to $J$. (b) The (black) filled circles
show the magnetization discontinuities, while the (red) filled squares the susceptibility discontinuities as functions of $\frac{J'}{J}$ and the magnetic field $h$. The (green) diamonds show the saturation field.
}
\label{fig:isolatedpentagonssubcluster}
\end{figure}

The importance of frustration is seen in the lowest energy configuration magnetic response of the group of six pentagons located at the ends of the molecules, when it is completely isolated (Fig. \ref{fig:isolatedpentagonssubcluster}(a)).
Pentagons introduce frustration in fullerene molecules, since an isolated pentagon can not have all antiferromagnetically interacting nearest-neighbor spin pairs simultaneously pointing in opposite directions in its lowest energy configuration, due to the odd number of spins and the closed periodic boundary conditions. In contrast, this is possible for an isolated hexagon. The magnetization curve of the six-pentagon subblock provides an insight on the properties of the whole cluster, as it is the sole carrier of frustration. Nearest neighbor intrapentagon interactions are taken equal to $J$ and interpentagon interactions equal to $J'$ (Fig. \ref{fig:isolatedpentagonssubcluster}(a)). The ground state magnetization and susceptibility discontinuities are shown in Fig. \ref{fig:isolatedpentagonssubcluster}(b) as functions of $\frac{J'}{J}$ and $h$. When $J'=0$ the magnetization curve corresponds to that of an isolated pentagon, which has constant susceptibility up to saturation. A non-zero $J'$ does not introduce further frustration as it does not change the lowest energy spin configuration of the six pentagons in zero field other than reducing its degeneracy, forcing neighboring spins belonging to different pentagons to be antiparallel. However the interpentagon coupling generates two ground state magnetization gaps in a field, which survive up to $J'=J$ (App. \ref{appendix:subblocksixnonneighboringpentagons}). The lowest field magnetization discontinuity is flanked by susceptibility discontinuities, which disappear for higher $J'$.


\subsection{$D_{5d}$ Symmetry Molecules}
\label{subsec:D5dsymmetrymolecules}

\begin{figure}
\includegraphics[width=3.5in,height=2.5in]{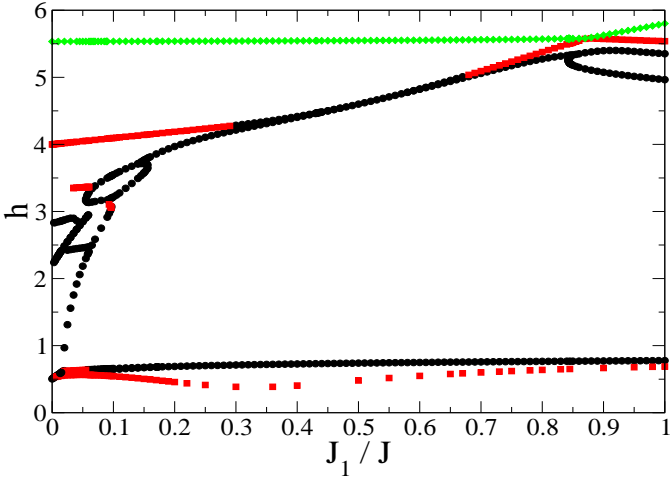}
\vspace{0pt}
\caption{(Color online) Evolution of the discontinuities for the $D_{5d}$ molecule with $N=80$ as a function of the relative coupling $\frac{J_1}{J}$ between the two six-pentagon subblocks at the edges and the (5,5) armchair nanotube, and the magnetic field $h$. The (black) filled circles show the magnetization discontinuities, while the (red) filled squares the susceptibility discontinuities. The (green) diamonds show the saturation field.
}
\label{fig:eightysiteslinking}
\end{figure}

The first class of molecules considered in this paper is formed by filling the space between two of the six pentagon subblocks of Sec. \ref{subsec:sixnonneighboringpentagons} with a (5,5) armchair carbon nanotube structure, which is solely made up of hexagons. When the pentagons and the nanotube structure are brought together to form the cluster further frustration is introduced, as the lowest energy is slightly higher than the sum of the energies of the isolated pentagons and hexagons (App. \ref{appendix:zerofieldgroundstatesaturationfield}). Furthermore it is found that the ground state magnetization curve of the fullerene molecules under consideration has more discontinuities than those of the isolated six pentagon subblocks, with the number of discontinuities increasing with $N$. The smallest member of the class is the truncated icosahedron for which $N=60$, which has the two caps directly linked and is already known to possess two magnetization jumps \cite{NPK07}. The smallest member of the class with $D_{5d}$ symmetry has $N=80$ and five discontinuities in total. Fig. \ref{fig:eightysiteslinking} shows how these discontinuities evolve from the isolated six-pentagon subblock limit when 20 spins are introduced in the middle to form the cluster. The coupling between the nanotube part and the pentagon subblocks equals $J_1$. When the three parts that form the molecule are not interacting ($J_1=0$) there are two magnetization discontinuities due to the pentagon subblocks according to Fig. \ref{fig:isolatedpentagonssubcluster}(b), as well as a susceptibility discontinuity originating in the saturation field of the nanotube part. This equals 4, which is the saturation field of an isolated hexagon. Increasing the coupling $J_1$ between the nanotube and the pentagon subblocks results in a rather complicated discontinuity diagram plotted as a function of $J_1$ over the intrapentagon subblock and intrananotube coupling $J$ and the magnetic field $h$, with many magnetization and susceptibility discontinuities. The low-field susceptibility and magnetization jumps at the uniform interaction limit $J_1=J$ are traced back to the low-field magnetization discontinuity of the pentagon subblocks. On the other hand, the high-field discontinuities at $J_1=J$ trace themselves back to all the discontinuities present when $J_1=0$. For the icosahedron it has been shown that its classical magnetization gap originates in the magnetization discontinuity of two isolated spins that connect with a triangular strip to form the icosahedron \cite{NPK15}. Figure \ref{fig:eightysiteslinking} shows that similar considerations for the $N=80$ cluster lead to more complicated results, as the parts that make it up are more complex in comparison with the parts that make up the icosahedron.



\begin{figure}
\includegraphics[width=3.5in,height=2.5in]{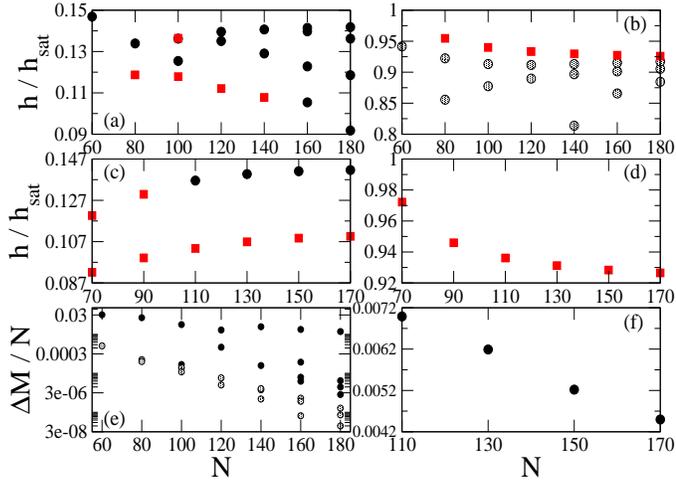}
\vspace{0pt}
\caption{(Color online) (a), (b) Magnetization and susceptibility discontinuities for the $D_{5d}$ molecules which have the shape of (5,5) armchair nanotubes and are capped on both sides with six pentagons not neighboring each other. They are given as functions of the number of vertices $N$ and the reduced magnetic field $\frac{h}{h_{sat}}$, with $h_{sat}$ the saturation magnetic field ($N=60$ corresponds to the $I_h$ symmetry molecule \cite{NPK07}). The (black) filled circles and the (black) dotted circles show respectively the low- and high-field magnetization discontinuities, while the (red) filled squares the susceptibility discontinuities. (c), (d) Similarly for the $D_{5h}$ molecules. (e), (f) Corresponding strengths of the magnetization discontinuities per spin $\frac{\Delta M}{N}$ for the $D_{5d}$ and $D_{5h}$ molecules respectively.
}
\label{fig:D5dD5hmagnsusc}
\end{figure}

Figures \ref{fig:D5dD5hmagnsusc}(a) and \ref{fig:D5dD5hmagnsusc}(b) show respectively the location of the low- and high-field ground state magnetization and susceptibility discontinuities for the $D_{5d}$ molecules as functions of $N$ and the reduced magnetic field $\frac{h}{h_{sat}}$, with $h_{sat}$ the saturation magnetic field (for the numerical values see App. \ref{appendix:groundstatechanges}).
For both low and high fields the number of magnetization jumps increases with $N$. Initially there is only a single low-field magnetization gap for $N=60$, and by $N=180$ there are already four. For high fields the corresponding numbers are one for $N=60$ and three for $N=180$, to which a susceptibility discontinuity is added. For the $N=180$ cluster there are already seven magnetization and one susceptibility jumps in total, a number far superseding the three magnetization gaps of the dodecahedron. Figure \ref{fig:D5dD5hmagnsusc}(a) hints to the fact that new ground state discontinuities emerge close to the highest low-field discontinuity, as can be seen for $N=120$ and 160. The discontinuities move away from each other with increasing $N$. Similarly for the high-field jumps in Fig. \ref{fig:D5dD5hmagnsusc}(b), new ones emerge for $N=80$ and 140, but now the discontinuities approach each other with $N$. The low-field susceptibility discontinuity gives way to a magnetization discontinuity for higher $N$, while the high-field susceptibility jump persists up to $N=180$. Another feature of the jumps at least for high fields is that they tend to constant values of $\frac{h}{h_{sat}}$ with $N$. The corresponding changes in the magnetization per spin are shown in Fig. \ref{fig:D5dD5hmagnsusc}(e) (for the numerical values see App. \ref{appendix:groundstatechanges}). The low-field ground state discontinuities are associated with a stronger magnetization change, which is as big as $\Delta M=0.760$ for $N=180$.

Specific features of the discontinuities can be detected (App. \ref{appendix:groundstatechanges}). For low fields as well as just before saturation some of the pentagon spins are parallel or antiparallel with the magnetic field, and their deviation from or alignment with these directions generates respectively the lowest and highest field discontinuities. The last low-field magnetization discontinuity, which is associated with the strongest magnetization change, occurs as pentagon spins change from having the lowest to having the highest total magnetizations along the field. After this discontinuity all spins have negative magnetic energy.
However inspection of the lowest energy configuration symmetries (App. \ref{appendix:symmetrygroundstate}) does not reveal a pattern away from small and high fields that could lead to the identification of the number of discontinuities for arbitrary $N$ without performing the full calculation. In addition the calculation is numerically demanding for a complete detection of the jumps for $N>180$, and as discussed earlier in this section Fig. \ref{fig:eightysiteslinking} shows that it is not straightforward to identify a simple origin of the discontinuities even for $N=80$, both of which would also be helpful for the identification of the number of gaps as a function of $N$.

\subsection{$D_{5h}$ Symmetry Molecules}
\label{subsec:D5hsymmetrymolecules}

The importance of the multiple discontinuities of the (5,5) armchair-type $D_{5d}$ molecules is highlighted by comparing with the corresponding number for the slightly different $D_{5h}$ molecules, plotted in Figs. \ref{fig:D5dD5hmagnsusc}(c) and \ref{fig:D5dD5hmagnsusc}(d). Their number is fixed for all clusters (up to $N=170$ was calculated), and the only change is that a low-field susceptibility gap becomes one of the magnetization for $N \geq 110$. Again they tend to approach constant values of $\frac{h}{h_{sat}}$ with $N$. In fact, the low-field magnetization discontinuity corresponds to the highest low-field magnetization jump of the $D_{5d}$ molecules, and the high-field susceptibility gap to the high-field susceptibility discontinuity of the $D_{5d}$ molecules. The strength of the magnetization discontinuity per spin is shown in Fig. \ref{fig:D5dD5hmagnsusc}(f), and equals $\Delta M=0.764$ for $N=170$. The different behavior of the $D_{5d}$ and $D_{5h}$ molecules
shows the importance of subtle spatial symmetry differences for the magnetic response. Even though both
resemble (5,5) armchair nanotubes and only differ in symmetry by having a center of inversion ($D_{5d}$) instead of a plane of symmetry ($D_{5h}$), their magnetic response is quite different.

\subsection{(5,0) Zigzag Nanotubes Capped with Neighboring Pentagons}
\label{subsec:zigzagnanotubescappedneighboringpentagons}

\begin{figure}
\centerline{
\includegraphics[width=1.6in,height=1.4in]{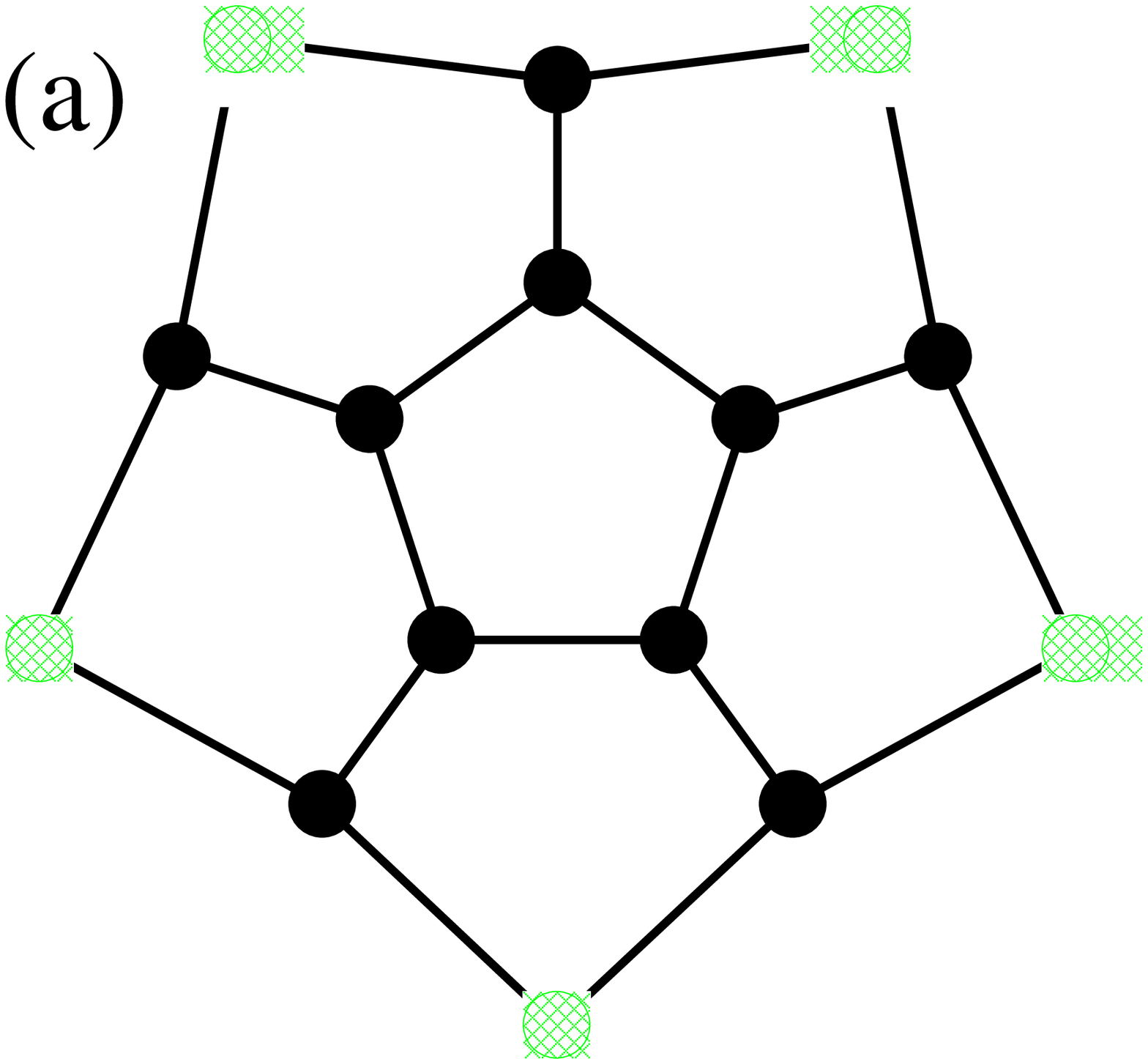}
\includegraphics[width=1.9in,height=1.4in]{Figure8}
}
\vspace{0pt}
\caption{(Color online) (a) Planar projection of the six non-isolated pentagons subblock that is the source of frustration for the $D_{5d}$ and $D_{5h}$ fullerene molecules resembling (5,0) zigzag nanotubes. The circles are classical spins of unit magnitude, with the (green) patterned circles connecting the subblock with the nanotube part of the molecule. Each spin interacts with its nearest neighbors according to the connecting lines, which all correspond to the same coupling strength. (b) Magnetization per spin $\frac{M}{N}$ as a function of the magnetic field $h$. The (red) arrows point to the two susceptibility discontinuities.
}
\label{fig:nonisolatedpentagonssubcluster}
\end{figure}

To investigate if magnetization patterns similar to the one of the (5,5) armchair nanotube-type molecules characterize other classes, the magnetization response of fullerene molecules that resemble (5,0) zigzag nanotubes was calculated. These are capped at their edges with six pentagons neighboring each other (Fig. \ref{fig:nonisolatedpentagonssubcluster}(a)).
Exactly like the molecules capped with isolated pentagons their number of vertices is a multiple of 10, and they have $D_{5d}$ symmetry when it is an even (Fig. \ref{fig:zigzag}(a)) and $D_{5h}$ symmetry when it is an odd multiple of 10 (Fig. \ref{fig:zigzag}(b)) \cite{Fowler95}. The magnetization response of the unit that causes frustration when isolated is shown in Fig. \ref{fig:nonisolatedpentagonssubcluster}(b). It has two susceptibility discontinuities, showing again that the pentagons are a source of discontinuous response in fullerene molecules, even though this time the discontinuities are not of the magnetization as in the (5,5) armchair-like case.

\begin{figure}
\includegraphics[width=3.5in,height=2.5in]{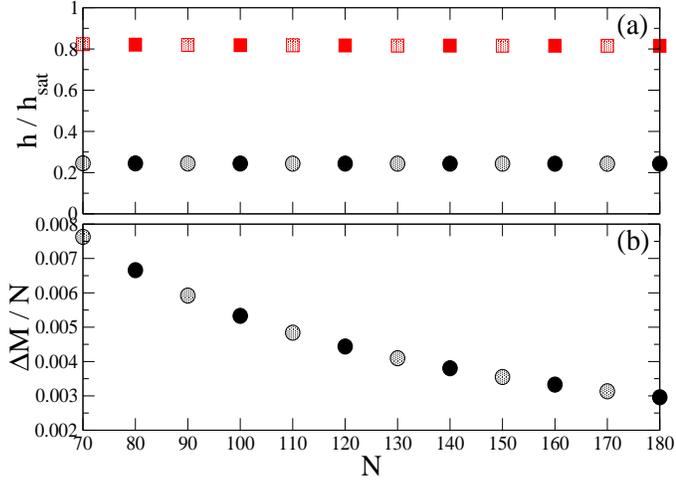}
\vspace{0pt}
\caption{(Color online) (a) Magnetization and susceptibility discontinuities for the $D_{5d}$ and $D_{5h}$ molecules which have the shape of (5,0) zigzag nanotubes and are capped on both sides with six pentagons neighboring each other. They are given as functions of the number of vertices $N$ and the reduced magnetic field $\frac{h}{h_{sat}}$, with $h_{sat}$ the saturation magnetic field. The (black) filled and the (black) dotted circles show respectively the magnetization discontinuities for the $D_{5d}$ and $D_{5h}$ molecules, while the (red) filled and the (red) dotted squares the corresponding susceptibility discontinuities. (c) Corresponding strength of the magnetization discontinuity per spin $\frac{\Delta M}{N}$.
}
\label{fig:D5dD5hnonisolatedmagnsusc}
\end{figure}

The locations of the magnetization and susceptibility discontinuities and the corresponding change in the magnetization per spin for this class of fullerene molecules are plotted in Fig. \ref{fig:D5dD5hnonisolatedmagnsusc}. In this case there is only one low-field magnetization and one high-field susceptibility discontinuity, whose locations with respect to $\frac{h}{h_{sat}}$ are practically constant already for relatively small $N$. In addition, spatial symmetry makes no difference in the magnetic response of the (5,0) zigzag-type molecules, unlike the (5,5) armchair-type molecules.

Since the (5,0) zigzag have a different cap than the (5,5) armchair nanotubes, the two cases are not directly comparable. The different chiralities require a different cap to form a fullerene molecule. Still it is interesting that for the armchair-type molecules subtle differences in the spatial symmetry make a significant difference in the magnetic response. More light on these issues can be shed by examining other nanotube-like fullerene molecules, however such a task is limited by numerical requirements as the molecules increase in size.



\section{Conclusions}
\label{sec:conclusions}

It was shown that a class of fullerene molecules of $D_{5d}$ spatial symmetry, which have the shape of (5,5) armchair carbon nanotubes and are capped at the edges with six non-neighboring pentagon subblocks, have a number of classical ground state magnetization discontinuities which increases with their size. This occurs when classical spins mounted at their vertices interact according to the AHM. The effect is
sensitive to spatial symmetry as for the very similar armchair-type $D_{5h}$ molecules the number of discontinuities does not change with their size. It is also sensitive to the chirality of the nanotube, with fullerene molecules based on (5,0) zigzag nanotubes and capped at the edges with six neighboring pentagon subblocks having a fixed number of discontinuities, which furthermore does not depend on their spatial symmetry.
Richer magnetic behavior is expected when other frustrated fullerene molecules are encapsulated in the ones considered in this paper. The findings in this paper show that the specific topology of the molecules causes unexpected magnetic response within the framework of the classical AHM, and it is hoped that in the quantum case relaxing the limit of large on-site interaction $U$ to more realistic values will also generate interesting magnetic behavior \cite{Yazyev10}.









\bibliography{paperfullerenes}

\begin{thebibliography}{84}
\expandafter\ifx\csname natexlab\endcsname\relax\def\natexlab#1{#1}\fi
\expandafter\ifx\csname bibnamefont\endcsname\relax
  \def\bibnamefont#1{#1}\fi
\expandafter\ifx\csname bibfnamefont\endcsname\relax
  \def\bibfnamefont#1{#1}\fi
\expandafter\ifx\csname citenamefont\endcsname\relax
  \def\citenamefont#1{#1}\fi
\expandafter\ifx\csname url\endcsname\relax
  \def\url#1{\texttt{#1}}\fi
\expandafter\ifx\csname urlprefix\endcsname\relax\def\urlprefix{URL }\fi
\providecommand{\bibinfo}[2]{#2}
\providecommand{\eprint}[2][]{\url{#2}}

\bibitem[{\citenamefont{Fowler and Manolopoulos}(1995)}]{Fowler95}
\bibinfo{author}{\bibfnamefont{P.~W.} \bibnamefont{Fowler}} \bibnamefont{and}
  \bibinfo{author}{\bibfnamefont{D.~E.} \bibnamefont{Manolopoulos}},
  \emph{\bibinfo{title}{An Atlas of Fullerenes}} (\bibinfo{publisher}{Oxford
  University Press}, \bibinfo{address}{Oxford}, \bibinfo{year}{1995}).

\bibitem[{\citenamefont{Anderson}(1973)}]{Anderson73}
\bibinfo{author}{\bibfnamefont{P.~W.} \bibnamefont{Anderson}},
  \bibinfo{journal}{Mater. Res. Bull.} \textbf{\bibinfo{volume}{8}},
  \bibinfo{pages}{153} (\bibinfo{year}{1973}).

\bibitem[{\citenamefont{Diep}(2005)}]{Diep05}
\bibinfo{author}{\bibfnamefont{H.~T.} \bibnamefont{Diep}},
  \emph{\bibinfo{title}{Frustrated Spin Systems}} (\bibinfo{publisher}{World
  Scientific}, \bibinfo{address}{Singapore}, \bibinfo{year}{2005}).

\bibitem[{\citenamefont{Schnack}(2010)}]{Schnack10}
\bibinfo{author}{\bibfnamefont{J.}~\bibnamefont{Schnack}},
  \bibinfo{journal}{Dalton Trans.} \textbf{\bibinfo{volume}{39}},
  \bibinfo{pages}{4677} (\bibinfo{year}{2010}).

\bibitem[{\citenamefont{Konstantinidis and Coffey}(2001)}]{NPK01}
\bibinfo{author}{\bibfnamefont{N.~P.} \bibnamefont{Konstantinidis}}
  \bibnamefont{and} \bibinfo{author}{\bibfnamefont{D.}~\bibnamefont{Coffey}},
  \bibinfo{journal}{Phys. Rev. B} \textbf{\bibinfo{volume}{63}},
  \bibinfo{pages}{184436} (\bibinfo{year}{2001}).

\bibitem[{\citenamefont{Konstantinidis and Coffey}(2002)}]{NPK02}
\bibinfo{author}{\bibfnamefont{N.~P.} \bibnamefont{Konstantinidis}}
  \bibnamefont{and} \bibinfo{author}{\bibfnamefont{D.}~\bibnamefont{Coffey}},
  \bibinfo{journal}{Phys. Rev. B} \textbf{\bibinfo{volume}{66}},
  \bibinfo{pages}{174426} (\bibinfo{year}{2002}).

\bibitem[{\citenamefont{Konstantinidis and Patterson}(2004)}]{NPK04}
\bibinfo{author}{\bibfnamefont{N.~P.} \bibnamefont{Konstantinidis}}
  \bibnamefont{and} \bibinfo{author}{\bibfnamefont{C.~H.}
  \bibnamefont{Patterson}}, \bibinfo{journal}{Phys. Rev. B}
  \textbf{\bibinfo{volume}{70}}, \bibinfo{pages}{064407}
  (\bibinfo{year}{2004}).

\bibitem[{\citenamefont{Konstantinidis and Lounis}(2013)}]{NPK13}
\bibinfo{author}{\bibfnamefont{N.~P.} \bibnamefont{Konstantinidis}}
  \bibnamefont{and} \bibinfo{author}{\bibfnamefont{S.}~\bibnamefont{Lounis}},
  \bibinfo{journal}{Phys. Rev. B} \textbf{\bibinfo{volume}{88}},
  \bibinfo{pages}{184414} (\bibinfo{year}{2013}).

\bibitem[{\citenamefont{Altmann and Herzig}(1994)}]{Altmann94}
\bibinfo{author}{\bibfnamefont{S.~L.} \bibnamefont{Altmann}} \bibnamefont{and}
  \bibinfo{author}{\bibfnamefont{P.}~\bibnamefont{Herzig}},
  \emph{\bibinfo{title}{Point-Group Theory Tables}} (\bibinfo{publisher}{Oxford
  University Press}, \bibinfo{address}{London}, \bibinfo{year}{1994}).

\bibitem[{\citenamefont{Hebard et~al.}(1991)\citenamefont{Hebard, Rosseinsky,
  Haddon, Murphy, Glarum, Palstra, Ramirez, and Kortan}}]{Hebard91}
\bibinfo{author}{\bibfnamefont{A.~F.} \bibnamefont{Hebard}},
  \bibinfo{author}{\bibfnamefont{M.~J.} \bibnamefont{Rosseinsky}},
  \bibinfo{author}{\bibfnamefont{R.~C.} \bibnamefont{Haddon}},
  \bibinfo{author}{\bibfnamefont{D.~W.} \bibnamefont{Murphy}},
  \bibinfo{author}{\bibfnamefont{S.~H.} \bibnamefont{Glarum}},
  \bibinfo{author}{\bibfnamefont{T.~T.~M.} \bibnamefont{Palstra}},
  \bibinfo{author}{\bibfnamefont{A.~P.} \bibnamefont{Ramirez}},
  \bibnamefont{and} \bibinfo{author}{\bibfnamefont{A.~R.}
  \bibnamefont{Kortan}}, \bibinfo{journal}{Nature (London)}
  \textbf{\bibinfo{volume}{350}}, \bibinfo{pages}{600} (\bibinfo{year}{1991}).

\bibitem[{\citenamefont{Gunnarsson}(1997)}]{Gunnarsson97}
\bibinfo{author}{\bibfnamefont{O.}~\bibnamefont{Gunnarsson}},
  \bibinfo{journal}{Rev. Mod. Phys.} \textbf{\bibinfo{volume}{69}},
  \bibinfo{pages}{575} (\bibinfo{year}{1997}).

\bibitem[{\citenamefont{Saito and Oshiyama}(1991)}]{Saito91}
\bibinfo{author}{\bibfnamefont{S.}~\bibnamefont{Saito}} \bibnamefont{and}
  \bibinfo{author}{\bibfnamefont{A.}~\bibnamefont{Oshiyama}},
  \bibinfo{journal}{Phys. Rev. B (R)} \textbf{\bibinfo{volume}{44}},
  \bibinfo{pages}{11536} (\bibinfo{year}{1991}).

\bibitem[{\citenamefont{Erwin and Pederson}(1991)}]{Erwin91}
\bibinfo{author}{\bibfnamefont{S.~C.} \bibnamefont{Erwin}} \bibnamefont{and}
  \bibinfo{author}{\bibfnamefont{M.~R.} \bibnamefont{Pederson}},
  \bibinfo{journal}{Phys. Rev. Lett.} \textbf{\bibinfo{volume}{67}},
  \bibinfo{pages}{1610} (\bibinfo{year}{1991}).

\bibitem[{\citenamefont{Kiefl et~al.}(1992)\citenamefont{Kiefl, Duty,
  Schneider, MacFarlane, Chow, Elzey, Mendels, Morris, Brewer, Ansaldo
  et~al.}}]{Kiefl92}
\bibinfo{author}{\bibfnamefont{R.~F.} \bibnamefont{Kiefl}},
  \bibinfo{author}{\bibfnamefont{T.~L.} \bibnamefont{Duty}},
  \bibinfo{author}{\bibfnamefont{J.~W.} \bibnamefont{Schneider}},
  \bibinfo{author}{\bibfnamefont{A.}~\bibnamefont{MacFarlane}},
  \bibinfo{author}{\bibfnamefont{K.}~\bibnamefont{Chow}},
  \bibinfo{author}{\bibfnamefont{J.}~\bibnamefont{Elzey}},
  \bibinfo{author}{\bibfnamefont{P.}~\bibnamefont{Mendels}},
  \bibinfo{author}{\bibfnamefont{G.~D.} \bibnamefont{Morris}},
  \bibinfo{author}{\bibfnamefont{J.~H.} \bibnamefont{Brewer}},
  \bibinfo{author}{\bibfnamefont{E.~J.} \bibnamefont{Ansaldo}},
  \bibnamefont{et~al.}, \bibinfo{journal}{Phys. Rev. Lett.}
  \textbf{\bibinfo{volume}{69}}, \bibinfo{pages}{2005} (\bibinfo{year}{1992}).

\bibitem[{\citenamefont{Murphy et~al.}(1992)\citenamefont{Murphy, Rosseinsky,
  Fleming, Tycko, Ramirez, Haddon, Siegrist, Dabbagh, Tully, and
  Walstedt}}]{Murphy92}
\bibinfo{author}{\bibfnamefont{D.~W.} \bibnamefont{Murphy}},
  \bibinfo{author}{\bibfnamefont{M.~J.} \bibnamefont{Rosseinsky}},
  \bibinfo{author}{\bibfnamefont{R.~M.} \bibnamefont{Fleming}},
  \bibinfo{author}{\bibfnamefont{R.}~\bibnamefont{Tycko}},
  \bibinfo{author}{\bibfnamefont{A.~P.} \bibnamefont{Ramirez}},
  \bibinfo{author}{\bibfnamefont{R.~C.} \bibnamefont{Haddon}},
  \bibinfo{author}{\bibfnamefont{T.}~\bibnamefont{Siegrist}},
  \bibinfo{author}{\bibfnamefont{G.}~\bibnamefont{Dabbagh}},
  \bibinfo{author}{\bibfnamefont{J.~C.} \bibnamefont{Tully}}, \bibnamefont{and}
  \bibinfo{author}{\bibfnamefont{R.~E.} \bibnamefont{Walstedt}},
  \bibinfo{journal}{J. Phys. Chem. Solids} \textbf{\bibinfo{volume}{53}},
  \bibinfo{pages}{1321} (\bibinfo{year}{1992}).

\bibitem[{\citenamefont{Benning et~al.}(1993)\citenamefont{Benning, Stepniak,
  Poirier, Martins, Weaver, Chibante, and Smalley}}]{Benning93}
\bibinfo{author}{\bibfnamefont{P.~J.} \bibnamefont{Benning}},
  \bibinfo{author}{\bibfnamefont{F.}~\bibnamefont{Stepniak}},
  \bibinfo{author}{\bibfnamefont{D.~M.} \bibnamefont{Poirier}},
  \bibinfo{author}{\bibfnamefont{J.~L.} \bibnamefont{Martins}},
  \bibinfo{author}{\bibfnamefont{J.~H.} \bibnamefont{Weaver}},
  \bibinfo{author}{\bibfnamefont{L.~P.~F.} \bibnamefont{Chibante}},
  \bibnamefont{and} \bibinfo{author}{\bibfnamefont{R.~E.}
  \bibnamefont{Smalley}}, \bibinfo{journal}{Phys. Rev. B}
  \textbf{\bibinfo{volume}{47}}, \bibinfo{pages}{13843} (\bibinfo{year}{1993}).

\bibitem[{\citenamefont{Yang et~al.}(2003)\citenamefont{Yang, Brouet, Zhou,
  Choi, Louie, Cohen, Kellar, Bogdanov, Lanzara, Goldoni et~al.}}]{Yang03}
\bibinfo{author}{\bibfnamefont{W.~L.} \bibnamefont{Yang}},
  \bibinfo{author}{\bibfnamefont{V.}~\bibnamefont{Brouet}},
  \bibinfo{author}{\bibfnamefont{X.~J.} \bibnamefont{Zhou}},
  \bibinfo{author}{\bibfnamefont{H.~J.} \bibnamefont{Choi}},
  \bibinfo{author}{\bibfnamefont{S.~G.} \bibnamefont{Louie}},
  \bibinfo{author}{\bibfnamefont{M.~L.} \bibnamefont{Cohen}},
  \bibinfo{author}{\bibfnamefont{S.~A.} \bibnamefont{Kellar}},
  \bibinfo{author}{\bibfnamefont{P.~V.} \bibnamefont{Bogdanov}},
  \bibinfo{author}{\bibfnamefont{A.}~\bibnamefont{Lanzara}},
  \bibinfo{author}{\bibfnamefont{A.}~\bibnamefont{Goldoni}},
  \bibnamefont{et~al.}, \bibinfo{journal}{Science}
  \textbf{\bibinfo{volume}{300}}, \bibinfo{pages}{303} (\bibinfo{year}{2003}).

\bibitem[{\citenamefont{Chakravarty et~al.}(1991)\citenamefont{Chakravarty,
  Gelfand, and Kivelson}}]{Chakravarty91}
\bibinfo{author}{\bibfnamefont{S.}~\bibnamefont{Chakravarty}},
  \bibinfo{author}{\bibfnamefont{M.}~\bibnamefont{Gelfand}}, \bibnamefont{and}
  \bibinfo{author}{\bibfnamefont{S.}~\bibnamefont{Kivelson}},
  \bibinfo{journal}{Science} \textbf{\bibinfo{volume}{254}},
  \bibinfo{pages}{970} (\bibinfo{year}{1991}).

\bibitem[{\citenamefont{Lin et~al.}(2007)\citenamefont{Lin, S{\o}rensen,
  Kallin, and Berlinsky}}]{Lin07}
\bibinfo{author}{\bibfnamefont{F.}~\bibnamefont{Lin}},
  \bibinfo{author}{\bibfnamefont{E.~S.} \bibnamefont{S{\o}rensen}},
  \bibinfo{author}{\bibfnamefont{C.}~\bibnamefont{Kallin}}, \bibnamefont{and}
  \bibinfo{author}{\bibfnamefont{A.~J.} \bibnamefont{Berlinsky}},
  \bibinfo{journal}{Phys. Rev. B} \textbf{\bibinfo{volume}{75}},
  \bibinfo{pages}{075112} (\bibinfo{year}{2007}).

\bibitem[{\citenamefont{Auerbach}(1998)}]{Auerbach98}
\bibinfo{author}{\bibfnamefont{A.}~\bibnamefont{Auerbach}},
  \emph{\bibinfo{title}{Interacting Electrons and Quantum Magnetism}}
  (\bibinfo{publisher}{Springer Verlag}, \bibinfo{address}{New York},
  \bibinfo{year}{1998}).

\bibitem[{\citenamefont{Fazekas}(1999)}]{Fazekas99}
\bibinfo{author}{\bibfnamefont{P.}~\bibnamefont{Fazekas}},
  \emph{\bibinfo{title}{Lecture Notes on Electron Correlation and Magnetism}}
  (\bibinfo{publisher}{World Scientific}, \bibinfo{address}{Singapore},
  \bibinfo{year}{1999}).

\bibitem[{\citenamefont{Coffey and Trugman}(1992)}]{Coffey92}
\bibinfo{author}{\bibfnamefont{D.}~\bibnamefont{Coffey}} \bibnamefont{and}
  \bibinfo{author}{\bibfnamefont{S.~A.} \bibnamefont{Trugman}},
  \bibinfo{journal}{Phys. Rev. Lett.} \textbf{\bibinfo{volume}{69}},
  \bibinfo{pages}{176} (\bibinfo{year}{1992}).

\bibitem[{\citenamefont{Stollhoff}(1991)}]{Stollhoff91}
\bibinfo{author}{\bibfnamefont{G.}~\bibnamefont{Stollhoff}},
  \bibinfo{journal}{Phys. Rev. B} \textbf{\bibinfo{volume}{44}},
  \bibinfo{pages}{10998} (\bibinfo{year}{1991}).

\bibitem[{\citenamefont{N\'eel}(1936)}]{Neel36}
\bibinfo{author}{\bibfnamefont{L.}~\bibnamefont{N\'eel}},
  \bibinfo{journal}{Ann. Phys. Paris} \textbf{\bibinfo{volume}{5}},
  \bibinfo{pages}{232} (\bibinfo{year}{1936}).

\bibitem[{\citenamefont{Konstantinidis}(2005)}]{NPK05}
\bibinfo{author}{\bibfnamefont{N.~P.} \bibnamefont{Konstantinidis}},
  \bibinfo{journal}{Phys. Rev. B} \textbf{\bibinfo{volume}{72}},
  \bibinfo{pages}{064453} (\bibinfo{year}{2005}).

\bibitem[{\citenamefont{Konstantinidis}(2007)}]{NPK07}
\bibinfo{author}{\bibfnamefont{N.~P.} \bibnamefont{Konstantinidis}},
  \bibinfo{journal}{Phys. Rev. B} \textbf{\bibinfo{volume}{76}},
  \bibinfo{pages}{104434} (\bibinfo{year}{2007}).

\bibitem[{\citenamefont{Konstantinidis}()}]{NPK17}
\bibinfo{author}{\bibfnamefont{N.~P.} \bibnamefont{Konstantinidis}},
  \emph{\bibinfo{title}{cond-mat/1702.06214}}.

\bibitem[{\citenamefont{Konstantinidis}(2009)}]{NPK09}
\bibinfo{author}{\bibfnamefont{N.~P.} \bibnamefont{Konstantinidis}},
  \bibinfo{journal}{Phys. Rev. B} \textbf{\bibinfo{volume}{80}},
  \bibinfo{pages}{134427} (\bibinfo{year}{2009}).

\bibitem[{\citenamefont{Schr{\"o}der et~al.}(2005)\citenamefont{Schr{\"o}der,
  Schmidt, Schnack, and Luban}}]{Schroeder05}
\bibinfo{author}{\bibfnamefont{C.}~\bibnamefont{Schr{\"o}der}},
  \bibinfo{author}{\bibfnamefont{H.-J.} \bibnamefont{Schmidt}},
  \bibinfo{author}{\bibfnamefont{J.}~\bibnamefont{Schnack}}, \bibnamefont{and}
  \bibinfo{author}{\bibfnamefont{M.}~\bibnamefont{Luban}},
  \bibinfo{journal}{Phys. Rev. Lett.} \textbf{\bibinfo{volume}{94}},
  \bibinfo{pages}{207203} (\bibinfo{year}{2005}).

\bibitem[{\citenamefont{Konstantinidis}(2015{\natexlab{a}})}]{NPK15}
\bibinfo{author}{\bibfnamefont{N.~P.} \bibnamefont{Konstantinidis}},
  \bibinfo{journal}{J. Phys.: Condens. Matter} \textbf{\bibinfo{volume}{27}},
  \bibinfo{pages}{076001} (\bibinfo{year}{2015}{\natexlab{a}}).

\bibitem[{\citenamefont{Konstantinidis}(2016{\natexlab{a}})}]{NPK16-1}
\bibinfo{author}{\bibfnamefont{N.~P.} \bibnamefont{Konstantinidis}},
  \bibinfo{journal}{J. Phys.: Condens. Matter} \textbf{\bibinfo{volume}{28}},
  \bibinfo{pages}{456003} (\bibinfo{year}{2016}{\natexlab{a}}).

\bibitem[{\citenamefont{Lin and S{\o}rensen}(2008)}]{Lin08}
\bibinfo{author}{\bibfnamefont{F.}~\bibnamefont{Lin}} \bibnamefont{and}
  \bibinfo{author}{\bibfnamefont{E.~S.} \bibnamefont{S{\o}rensen}},
  \bibinfo{journal}{Phys. Rev. B} \textbf{\bibinfo{volume}{78}},
  \bibinfo{pages}{085435} (\bibinfo{year}{2008}).

\bibitem[{\citenamefont{Iijima}(1991)}]{Iijima91}
\bibinfo{author}{\bibfnamefont{S.}~\bibnamefont{Iijima}},
  \bibinfo{journal}{Nature (London)} \textbf{\bibinfo{volume}{354}},
  \bibinfo{pages}{56} (\bibinfo{year}{1991}).

\bibitem[{\citenamefont{Saito et~al.}(1998)\citenamefont{Saito, Dresselhaus,
  and Dresselhaus}}]{Saito98}
\bibinfo{author}{\bibfnamefont{R.}~\bibnamefont{Saito}},
  \bibinfo{author}{\bibfnamefont{G.}~\bibnamefont{Dresselhaus}},
  \bibnamefont{and} \bibinfo{author}{\bibfnamefont{M.~S.}
  \bibnamefont{Dresselhaus}}, \emph{\bibinfo{title}{Physical Properties of
  Carbon Nanotubes}} (\bibinfo{publisher}{Imperial College Press},
  \bibinfo{address}{London}, \bibinfo{year}{1998}).

\bibitem[{\citenamefont{Dekker}(1999)}]{Dekker99}
\bibinfo{author}{\bibfnamefont{C.}~\bibnamefont{Dekker}},
  \bibinfo{journal}{Physics Today} \textbf{\bibinfo{volume}{52}},
  \bibinfo{pages}{22} (\bibinfo{year}{1999}).

\bibitem[{\citenamefont{Terrones}(2003)}]{Terrones03}
\bibinfo{author}{\bibfnamefont{M.}~\bibnamefont{Terrones}},
  \bibinfo{journal}{Annu. Rev. Mater. Res.} \textbf{\bibinfo{volume}{33}},
  \bibinfo{pages}{419} (\bibinfo{year}{2003}).

\bibitem[{\citenamefont{Volder et~al.}(2013)\citenamefont{Volder, Tawfick,
  Baughman, and Hart}}]{DeVolder13}
\bibinfo{author}{\bibfnamefont{M.~F. L.~D.} \bibnamefont{Volder}},
  \bibinfo{author}{\bibfnamefont{S.~H.} \bibnamefont{Tawfick}},
  \bibinfo{author}{\bibfnamefont{R.~H.} \bibnamefont{Baughman}},
  \bibnamefont{and} \bibinfo{author}{\bibfnamefont{A.~J.} \bibnamefont{Hart}},
  \bibinfo{journal}{Science} \textbf{\bibinfo{volume}{339}},
  \bibinfo{pages}{535} (\bibinfo{year}{2013}).

\bibitem[{\citenamefont{Peng et~al.}(2008)\citenamefont{Peng, Locascio, Zapol,
  Li, Mielke, Schatz, and Espinosa}}]{Peng08}
\bibinfo{author}{\bibfnamefont{B.}~\bibnamefont{Peng}},
  \bibinfo{author}{\bibfnamefont{M.}~\bibnamefont{Locascio}},
  \bibinfo{author}{\bibfnamefont{P.}~\bibnamefont{Zapol}},
  \bibinfo{author}{\bibfnamefont{S.}~\bibnamefont{Li}},
  \bibinfo{author}{\bibfnamefont{S.~L.} \bibnamefont{Mielke}},
  \bibinfo{author}{\bibfnamefont{G.~C.} \bibnamefont{Schatz}},
  \bibnamefont{and} \bibinfo{author}{\bibfnamefont{H.~D.}
  \bibnamefont{Espinosa}}, \bibinfo{journal}{Nat. Nanotechnol.}
  \textbf{\bibinfo{volume}{3}}, \bibinfo{pages}{626} (\bibinfo{year}{2008}).

\bibitem[{\citenamefont{Pop et~al.}(2006)\citenamefont{Pop, Mann, Wang,
  Goodson, and Dai}}]{Pop06}
\bibinfo{author}{\bibfnamefont{E.}~\bibnamefont{Pop}},
  \bibinfo{author}{\bibfnamefont{D.}~\bibnamefont{Mann}},
  \bibinfo{author}{\bibfnamefont{Q.}~\bibnamefont{Wang}},
  \bibinfo{author}{\bibfnamefont{K.}~\bibnamefont{Goodson}}, \bibnamefont{and}
  \bibinfo{author}{\bibfnamefont{H.~J.} \bibnamefont{Dai}},
  \bibinfo{journal}{Nano Lett.} \textbf{\bibinfo{volume}{6}},
  \bibinfo{pages}{96} (\bibinfo{year}{2006}).

\bibitem[{\citenamefont{Hayashi et~al.}(2003)\citenamefont{Hayashi, Kim,
  Matoba, Esaka, Nishimura, Tsukada, Endo, and Dresselhaus}}]{Hayashi03}
\bibinfo{author}{\bibfnamefont{T.}~\bibnamefont{Hayashi}},
  \bibinfo{author}{\bibfnamefont{Y.}~\bibnamefont{Kim}},
  \bibinfo{author}{\bibfnamefont{T.}~\bibnamefont{Matoba}},
  \bibinfo{author}{\bibfnamefont{M.}~\bibnamefont{Esaka}},
  \bibinfo{author}{\bibfnamefont{K.}~\bibnamefont{Nishimura}},
  \bibinfo{author}{\bibfnamefont{T.}~\bibnamefont{Tsukada}},
  \bibinfo{author}{\bibfnamefont{M.}~\bibnamefont{Endo}}, \bibnamefont{and}
  \bibinfo{author}{\bibfnamefont{M.~S.} \bibnamefont{Dresselhaus}},
  \bibinfo{journal}{Nano Lett.} \textbf{\bibinfo{volume}{3}},
  \bibinfo{pages}{887} (\bibinfo{year}{2003}).

\bibitem[{\citenamefont{de~Jonge et~al.}(2005)\citenamefont{de~Jonge,
  Doytcheva, Allioux, Kaiser, Mentink, Teo, Lacerda, and Milne}}]{deJonge05}
\bibinfo{author}{\bibfnamefont{N.}~\bibnamefont{de~Jonge}},
  \bibinfo{author}{\bibfnamefont{M.}~\bibnamefont{Doytcheva}},
  \bibinfo{author}{\bibfnamefont{M.}~\bibnamefont{Allioux}},
  \bibinfo{author}{\bibfnamefont{M.}~\bibnamefont{Kaiser}},
  \bibinfo{author}{\bibfnamefont{S.~A.~M.} \bibnamefont{Mentink}},
  \bibinfo{author}{\bibfnamefont{K.~B.~K.} \bibnamefont{Teo}},
  \bibinfo{author}{\bibfnamefont{R.~G.} \bibnamefont{Lacerda}},
  \bibnamefont{and} \bibinfo{author}{\bibfnamefont{W.~I.} \bibnamefont{Milne}},
  \bibinfo{journal}{Adv. Mater.} \textbf{\bibinfo{volume}{17}},
  \bibinfo{pages}{451} (\bibinfo{year}{2005}).

\bibitem[{\citenamefont{Wei et~al.}(2001)\citenamefont{Wei, Vajtai, and
  Ajayan}}]{Wei01}
\bibinfo{author}{\bibfnamefont{B.~Q.} \bibnamefont{Wei}},
  \bibinfo{author}{\bibfnamefont{R.}~\bibnamefont{Vajtai}}, \bibnamefont{and}
  \bibinfo{author}{\bibfnamefont{P.~M.} \bibnamefont{Ajayan}},
  \bibinfo{journal}{Appl. Phys. Lett.} \textbf{\bibinfo{volume}{79}},
  \bibinfo{pages}{1172} (\bibinfo{year}{2001}).

\bibitem[{\citenamefont{Mintmire et~al.}(1992)\citenamefont{Mintmire, Dunlap,
  and White}}]{Mintmire92}
\bibinfo{author}{\bibfnamefont{J.~W.} \bibnamefont{Mintmire}},
  \bibinfo{author}{\bibfnamefont{B.~I.} \bibnamefont{Dunlap}},
  \bibnamefont{and} \bibinfo{author}{\bibfnamefont{C.~T.} \bibnamefont{White}},
  \bibinfo{journal}{Phys. Rev. Lett.} \textbf{\bibinfo{volume}{68}},
  \bibinfo{pages}{631} (\bibinfo{year}{1992}).

\bibitem[{\citenamefont{Hamada et~al.}(1992)\citenamefont{Hamada, Sawada, and
  Oshiyama}}]{Hamada92}
\bibinfo{author}{\bibfnamefont{N.}~\bibnamefont{Hamada}},
  \bibinfo{author}{\bibfnamefont{S.-I.} \bibnamefont{Sawada}},
  \bibnamefont{and} \bibinfo{author}{\bibfnamefont{A.}~\bibnamefont{Oshiyama}},
  \bibinfo{journal}{Phys. Rev. Lett.} \textbf{\bibinfo{volume}{68}},
  \bibinfo{pages}{1579} (\bibinfo{year}{1992}).

\bibitem[{\citenamefont{Saito et~al.}(1992)\citenamefont{Saito, Fujita,
  Dresselhaus, and Dresselhaus}}]{Saito92}
\bibinfo{author}{\bibfnamefont{R.}~\bibnamefont{Saito}},
  \bibinfo{author}{\bibfnamefont{M.}~\bibnamefont{Fujita}},
  \bibinfo{author}{\bibfnamefont{G.}~\bibnamefont{Dresselhaus}},
  \bibnamefont{and} \bibinfo{author}{\bibfnamefont{M.~S.}
  \bibnamefont{Dresselhaus}}, \bibinfo{journal}{Appl. Phys. Lett.}
  \textbf{\bibinfo{volume}{60}}, \bibinfo{pages}{2204} (\bibinfo{year}{1992}).

\bibitem[{\citenamefont{Liang et~al.}(2001)\citenamefont{Liang, Bockrath,
  Bozovic, Hafner, Tinkham, and Park}}]{Liang01}
\bibinfo{author}{\bibfnamefont{W.}~\bibnamefont{Liang}},
  \bibinfo{author}{\bibfnamefont{M.}~\bibnamefont{Bockrath}},
  \bibinfo{author}{\bibfnamefont{D.}~\bibnamefont{Bozovic}},
  \bibinfo{author}{\bibfnamefont{J.~H.} \bibnamefont{Hafner}},
  \bibinfo{author}{\bibfnamefont{M.}~\bibnamefont{Tinkham}}, \bibnamefont{and}
  \bibinfo{author}{\bibfnamefont{H.}~\bibnamefont{Park}},
  \bibinfo{journal}{Nature} \textbf{\bibinfo{volume}{411}},
  \bibinfo{pages}{665} (\bibinfo{year}{2001}).

\bibitem[{\citenamefont{Tans et~al.}(1997)\citenamefont{Tans, Devoret, Dai,
  Thess, Smalley, Geerligs, and Dekker}}]{Tans97}
\bibinfo{author}{\bibfnamefont{S.~J.} \bibnamefont{Tans}},
  \bibinfo{author}{\bibfnamefont{M.~H.} \bibnamefont{Devoret}},
  \bibinfo{author}{\bibfnamefont{H.}~\bibnamefont{Dai}},
  \bibinfo{author}{\bibfnamefont{A.}~\bibnamefont{Thess}},
  \bibinfo{author}{\bibfnamefont{R.~E.} \bibnamefont{Smalley}},
  \bibinfo{author}{\bibfnamefont{L.~J.} \bibnamefont{Geerligs}},
  \bibnamefont{and} \bibinfo{author}{\bibfnamefont{C.}~\bibnamefont{Dekker}},
  \bibinfo{journal}{Nature} \textbf{\bibinfo{volume}{386}},
  \bibinfo{pages}{474} (\bibinfo{year}{1997}).

\bibitem[{\citenamefont{Bockrath et~al.}(1997)\citenamefont{Bockrath, Cobden,
  McEuen, Chopra, Zettl, Thess, and Smalley}}]{Bockrath97}
\bibinfo{author}{\bibfnamefont{M.}~\bibnamefont{Bockrath}},
  \bibinfo{author}{\bibfnamefont{D.~H.} \bibnamefont{Cobden}},
  \bibinfo{author}{\bibfnamefont{P.~L.} \bibnamefont{McEuen}},
  \bibinfo{author}{\bibfnamefont{N.~G.} \bibnamefont{Chopra}},
  \bibinfo{author}{\bibfnamefont{A.}~\bibnamefont{Zettl}},
  \bibinfo{author}{\bibfnamefont{A.}~\bibnamefont{Thess}}, \bibnamefont{and}
  \bibinfo{author}{\bibfnamefont{R.~E.} \bibnamefont{Smalley}},
  \bibinfo{journal}{Science} \textbf{\bibinfo{volume}{275}},
  \bibinfo{pages}{1922} (\bibinfo{year}{1997}).

\bibitem[{\citenamefont{Cobden and Nyg{\aa}rd}(2002)}]{Cobden02}
\bibinfo{author}{\bibfnamefont{D.~H.} \bibnamefont{Cobden}} \bibnamefont{and}
  \bibinfo{author}{\bibfnamefont{J.}~\bibnamefont{Nyg{\aa}rd}},
  \bibinfo{journal}{Phys. Rev. Lett.} \textbf{\bibinfo{volume}{89}},
  \bibinfo{pages}{046803} (\bibinfo{year}{2002}).

\bibitem[{\citenamefont{Hagiwara et~al.}(2014)\citenamefont{Hagiwara, Ikeda,
  Kida, Matsuda, Tadera, Kyakuno, Yanagi, Maniwa, and Okunishi}}]{Hagiwara14}
\bibinfo{author}{\bibfnamefont{M.}~\bibnamefont{Hagiwara}},
  \bibinfo{author}{\bibfnamefont{M.}~\bibnamefont{Ikeda}},
  \bibinfo{author}{\bibfnamefont{T.}~\bibnamefont{Kida}},
  \bibinfo{author}{\bibfnamefont{K.}~\bibnamefont{Matsuda}},
  \bibinfo{author}{\bibfnamefont{S.}~\bibnamefont{Tadera}},
  \bibinfo{author}{\bibfnamefont{H.}~\bibnamefont{Kyakuno}},
  \bibinfo{author}{\bibfnamefont{K.}~\bibnamefont{Yanagi}},
  \bibinfo{author}{\bibfnamefont{Y.}~\bibnamefont{Maniwa}}, \bibnamefont{and}
  \bibinfo{author}{\bibfnamefont{K.}~\bibnamefont{Okunishi}},
  \bibinfo{journal}{J. Phys. Soc. Jpn.} \textbf{\bibinfo{volume}{83}},
  \bibinfo{pages}{113706} (\bibinfo{year}{2014}).

\bibitem[{\citenamefont{Singh}(2013)}]{Singh13}
\bibinfo{author}{\bibfnamefont{R.}~\bibnamefont{Singh}}, \bibinfo{journal}{J.
  Magn. Magn. Mater.} \textbf{\bibinfo{volume}{346}}, \bibinfo{pages}{58}
  (\bibinfo{year}{2013}).

\bibitem[{\citenamefont{Zhang et~al.}(2011)\citenamefont{Zhang, Liu, Qin, and
  Hu}}]{Zhang11}
\bibinfo{author}{\bibfnamefont{Y.}~\bibnamefont{Zhang}},
  \bibinfo{author}{\bibfnamefont{H.}~\bibnamefont{Liu}},
  \bibinfo{author}{\bibfnamefont{H.}~\bibnamefont{Qin}}, \bibnamefont{and}
  \bibinfo{author}{\bibfnamefont{J.}~\bibnamefont{Hu}}, \bibinfo{journal}{Appl.
  Surf. Sci.} \textbf{\bibinfo{volume}{257}}, \bibinfo{pages}{10206}
  (\bibinfo{year}{2011}).

\bibitem[{\citenamefont{Zhou et~al.}(2009)\citenamefont{Zhou, Wang, Sun, Chen,
  Kawazoe, and Jena}}]{Zhou09}
\bibinfo{author}{\bibfnamefont{J.}~\bibnamefont{Zhou}},
  \bibinfo{author}{\bibfnamefont{Q.}~\bibnamefont{Wang}},
  \bibinfo{author}{\bibfnamefont{Q.}~\bibnamefont{Sun}},
  \bibinfo{author}{\bibfnamefont{X.~S.} \bibnamefont{Chen}},
  \bibinfo{author}{\bibfnamefont{Y.}~\bibnamefont{Kawazoe}}, \bibnamefont{and}
  \bibinfo{author}{\bibfnamefont{P.}~\bibnamefont{Jena}},
  \bibinfo{journal}{Nano Lett.} \textbf{\bibinfo{volume}{9}},
  \bibinfo{pages}{3867} (\bibinfo{year}{2009}).

\bibitem[{\citenamefont{Zhang and Guo}(2009)}]{Zhang09}
\bibinfo{author}{\bibfnamefont{Z.}~\bibnamefont{Zhang}} \bibnamefont{and}
  \bibinfo{author}{\bibfnamefont{W.}~\bibnamefont{Guo}}, \bibinfo{journal}{J.
  Am. Chem. Soc.} \textbf{\bibinfo{volume}{131}}, \bibinfo{pages}{6874}
  (\bibinfo{year}{2009}).

\bibitem[{\citenamefont{Ma et~al.}(2005)\citenamefont{Ma, Lehtinen, Foster, and
  Nieminen}}]{Ma05}
\bibinfo{author}{\bibfnamefont{Y.}~\bibnamefont{Ma}},
  \bibinfo{author}{\bibfnamefont{P.~O.} \bibnamefont{Lehtinen}},
  \bibinfo{author}{\bibfnamefont{A.~S.} \bibnamefont{Foster}},
  \bibnamefont{and} \bibinfo{author}{\bibfnamefont{R.~M.}
  \bibnamefont{Nieminen}}, \bibinfo{journal}{Phys. Rev. B}
  \textbf{\bibinfo{volume}{72}}, \bibinfo{pages}{085451}
  (\bibinfo{year}{2005}).

\bibitem[{\citenamefont{Moradian and Fathalian}(2006)}]{Moradian06}
\bibinfo{author}{\bibfnamefont{R.}~\bibnamefont{Moradian}} \bibnamefont{and}
  \bibinfo{author}{\bibfnamefont{A.}~\bibnamefont{Fathalian}},
  \bibinfo{journal}{Nanotechnology} \textbf{\bibinfo{volume}{17}},
  \bibinfo{pages}{1835} (\bibinfo{year}{2006}).

\bibitem[{\citenamefont{C{\'e}spedes et~al.}(2004)\citenamefont{C{\'e}spedes,
  Ferreira, Sanvito, Kociak, and Coey}}]{Cespedes04}
\bibinfo{author}{\bibfnamefont{O.}~\bibnamefont{C{\'e}spedes}},
  \bibinfo{author}{\bibfnamefont{M.~S.} \bibnamefont{Ferreira}},
  \bibinfo{author}{\bibfnamefont{S.}~\bibnamefont{Sanvito}},
  \bibinfo{author}{\bibfnamefont{M.}~\bibnamefont{Kociak}}, \bibnamefont{and}
  \bibinfo{author}{\bibfnamefont{J.~M.~D.} \bibnamefont{Coey}},
  \bibinfo{journal}{J. Phys.: Condens. Matter} \textbf{\bibinfo{volume}{16}},
  \bibinfo{pages}{L155} (\bibinfo{year}{2004}).

\bibitem[{\citenamefont{Geim and Novoselov}(2007)}]{Geim07}
\bibinfo{author}{\bibfnamefont{A.~K.} \bibnamefont{Geim}} \bibnamefont{and}
  \bibinfo{author}{\bibfnamefont{K.~S.} \bibnamefont{Novoselov}},
  \bibinfo{journal}{Nat. Mater.} \textbf{\bibinfo{volume}{6}},
  \bibinfo{pages}{183} (\bibinfo{year}{2007}).

\bibitem[{\citenamefont{Neto et~al.}(2009)\citenamefont{Neto, Guinea, Peres,
  Novoselov, and Geim}}]{Neto09}
\bibinfo{author}{\bibfnamefont{A.~H.~C.} \bibnamefont{Neto}},
  \bibinfo{author}{\bibfnamefont{F.}~\bibnamefont{Guinea}},
  \bibinfo{author}{\bibfnamefont{N.~M.~R.} \bibnamefont{Peres}},
  \bibinfo{author}{\bibfnamefont{K.~S.} \bibnamefont{Novoselov}},
  \bibnamefont{and} \bibinfo{author}{\bibfnamefont{A.~K.} \bibnamefont{Geim}},
  \bibinfo{journal}{Rev. Mod. Phys.} \textbf{\bibinfo{volume}{81}},
  \bibinfo{pages}{109} (\bibinfo{year}{2009}).

\bibitem[{\citenamefont{Yazyev}(2010)}]{Yazyev10}
\bibinfo{author}{\bibfnamefont{O.~V.} \bibnamefont{Yazyev}},
  \bibinfo{journal}{Rep. Prog. Phys.} \textbf{\bibinfo{volume}{73}},
  \bibinfo{pages}{056501} (\bibinfo{year}{2010}).

\bibitem[{\citenamefont{Deshpande et~al.}(2009)\citenamefont{Deshpande,
  Chandra, Caldwell, Novikov, Hone, and Bockrath}}]{Deshpande09}
\bibinfo{author}{\bibfnamefont{V.~V.} \bibnamefont{Deshpande}},
  \bibinfo{author}{\bibfnamefont{B.}~\bibnamefont{Chandra}},
  \bibinfo{author}{\bibfnamefont{R.}~\bibnamefont{Caldwell}},
  \bibinfo{author}{\bibfnamefont{D.~S.} \bibnamefont{Novikov}},
  \bibinfo{author}{\bibfnamefont{J.}~\bibnamefont{Hone}}, \bibnamefont{and}
  \bibinfo{author}{\bibfnamefont{M.}~\bibnamefont{Bockrath}},
  \bibinfo{journal}{Science} \textbf{\bibinfo{volume}{323}},
  \bibinfo{pages}{106} (\bibinfo{year}{2009}).

\bibitem[{\citenamefont{Bostwick et~al.}(2007)\citenamefont{Bostwick, Ohta,
  Seyller, Horn, and Rotenberg}}]{Bostwick07}
\bibinfo{author}{\bibfnamefont{A.}~\bibnamefont{Bostwick}},
  \bibinfo{author}{\bibfnamefont{T.}~\bibnamefont{Ohta}},
  \bibinfo{author}{\bibfnamefont{T.}~\bibnamefont{Seyller}},
  \bibinfo{author}{\bibfnamefont{K.}~\bibnamefont{Horn}}, \bibnamefont{and}
  \bibinfo{author}{\bibfnamefont{E.}~\bibnamefont{Rotenberg}},
  \bibinfo{journal}{Nat. Phys.} \textbf{\bibinfo{volume}{3}},
  \bibinfo{pages}{36} (\bibinfo{year}{2007}).

\bibitem[{\citenamefont{Kane et~al.}(1997)\citenamefont{Kane, Balents, and
  Fisher}}]{Kane97}
\bibinfo{author}{\bibfnamefont{C.}~\bibnamefont{Kane}},
  \bibinfo{author}{\bibfnamefont{L.}~\bibnamefont{Balents}}, \bibnamefont{and}
  \bibinfo{author}{\bibfnamefont{M.~P.~A.} \bibnamefont{Fisher}},
  \bibinfo{journal}{Phys. Rev. Lett.} \textbf{\bibinfo{volume}{79}},
  \bibinfo{pages}{5086} (\bibinfo{year}{1997}).

\bibitem[{\citenamefont{Bockrath et~al.}(1999)\citenamefont{Bockrath, Cobden,
  Lu, Rinzler, R.~E.~Smalley, and McEuen}}]{Bockrath99}
\bibinfo{author}{\bibfnamefont{M.}~\bibnamefont{Bockrath}},
  \bibinfo{author}{\bibfnamefont{D.~H.} \bibnamefont{Cobden}},
  \bibinfo{author}{\bibfnamefont{J.}~\bibnamefont{Lu}},
  \bibinfo{author}{\bibfnamefont{A.~G.} \bibnamefont{Rinzler}},
  \bibinfo{author}{\bibfnamefont{L.~B.} \bibnamefont{R.~E.~Smalley}},
  \bibnamefont{and} \bibinfo{author}{\bibfnamefont{P.~L.}
  \bibnamefont{McEuen}}, \bibinfo{journal}{Nature}
  \textbf{\bibinfo{volume}{397}}, \bibinfo{pages}{598} (\bibinfo{year}{1999}).

\bibitem[{\citenamefont{Siegel et~al.}(2013)\citenamefont{Siegel, Regan,
  Fedorov, Zettl, and Lanzara}}]{Siegel13}
\bibinfo{author}{\bibfnamefont{D.~A.} \bibnamefont{Siegel}},
  \bibinfo{author}{\bibfnamefont{W.}~\bibnamefont{Regan}},
  \bibinfo{author}{\bibfnamefont{A.~V.} \bibnamefont{Fedorov}},
  \bibinfo{author}{\bibfnamefont{A.}~\bibnamefont{Zettl}}, \bibnamefont{and}
  \bibinfo{author}{\bibfnamefont{A.}~\bibnamefont{Lanzara}},
  \bibinfo{journal}{Phys. Rev. Lett.} \textbf{\bibinfo{volume}{110}},
  \bibinfo{pages}{146802} (\bibinfo{year}{2013}).

\bibitem[{\citenamefont{Freitag et~al.}(2012)\citenamefont{Freitag, Trbovic,
  Weiss, and Sch{\"o}nenberger}}]{Freitag12}
\bibinfo{author}{\bibfnamefont{F.}~\bibnamefont{Freitag}},
  \bibinfo{author}{\bibfnamefont{J.}~\bibnamefont{Trbovic}},
  \bibinfo{author}{\bibfnamefont{M.}~\bibnamefont{Weiss}}, \bibnamefont{and}
  \bibinfo{author}{\bibfnamefont{C.}~\bibnamefont{Sch{\"o}nenberger}},
  \bibinfo{journal}{Phys. Rev. Lett.} \textbf{\bibinfo{volume}{108}},
  \bibinfo{pages}{076602} (\bibinfo{year}{2012}).

\bibitem[{\citenamefont{Zhou et~al.}(2008)\citenamefont{Zhou, Siegel, Fedorov,
  and Lanzara}}]{Zhou08}
\bibinfo{author}{\bibfnamefont{S.~Y.} \bibnamefont{Zhou}},
  \bibinfo{author}{\bibfnamefont{D.~A.} \bibnamefont{Siegel}},
  \bibinfo{author}{\bibfnamefont{A.~V.} \bibnamefont{Fedorov}},
  \bibnamefont{and} \bibinfo{author}{\bibfnamefont{A.}~\bibnamefont{Lanzara}},
  \bibinfo{journal}{Phys. Rev. Lett.} \textbf{\bibinfo{volume}{101}},
  \bibinfo{pages}{086402} (\bibinfo{year}{2008}).

\bibitem[{\citenamefont{Mayorov et~al.}(2011)\citenamefont{Mayorov, Elias,
  Mucha-Kruczynski, Gorbachev, Tudorovskiy, Zhukov, Morozov, Katsnelson, Falko,
  Geim et~al.}}]{Mayorov11}
\bibinfo{author}{\bibfnamefont{A.~S.} \bibnamefont{Mayorov}},
  \bibinfo{author}{\bibfnamefont{D.~C.} \bibnamefont{Elias}},
  \bibinfo{author}{\bibfnamefont{M.}~\bibnamefont{Mucha-Kruczynski}},
  \bibinfo{author}{\bibfnamefont{R.~V.} \bibnamefont{Gorbachev}},
  \bibinfo{author}{\bibfnamefont{T.}~\bibnamefont{Tudorovskiy}},
  \bibinfo{author}{\bibfnamefont{A.}~\bibnamefont{Zhukov}},
  \bibinfo{author}{\bibfnamefont{S.~V.} \bibnamefont{Morozov}},
  \bibinfo{author}{\bibfnamefont{M.~I.} \bibnamefont{Katsnelson}},
  \bibinfo{author}{\bibfnamefont{V.~I.} \bibnamefont{Falko}},
  \bibinfo{author}{\bibfnamefont{A.~K.} \bibnamefont{Geim}},
  \bibnamefont{et~al.}, \bibinfo{journal}{Science}
  \textbf{\bibinfo{volume}{333}}, \bibinfo{pages}{860} (\bibinfo{year}{2011}).

\bibitem[{\citenamefont{Luu and L{\"a}hde}(2016)}]{Luu16}
\bibinfo{author}{\bibfnamefont{T.}~\bibnamefont{Luu}} \bibnamefont{and}
  \bibinfo{author}{\bibfnamefont{T.~A.} \bibnamefont{L{\"a}hde}},
  \bibinfo{journal}{Phys. Rev. B} \textbf{\bibinfo{volume}{93}},
  \bibinfo{pages}{155106} (\bibinfo{year}{2016}).

\bibitem[{\citenamefont{Egger and Gogolin}(1997)}]{Egger97}
\bibinfo{author}{\bibfnamefont{R.}~\bibnamefont{Egger}} \bibnamefont{and}
  \bibinfo{author}{\bibfnamefont{A.~O.} \bibnamefont{Gogolin}},
  \bibinfo{journal}{Phys. Rev. Lett.} \textbf{\bibinfo{volume}{79}},
  \bibinfo{pages}{5082} (\bibinfo{year}{1997}).

\bibitem[{\citenamefont{Lin et~al.}(1998)\citenamefont{Lin, Balents, and
  Fisher}}]{Lin98}
\bibinfo{author}{\bibfnamefont{H.-H.} \bibnamefont{Lin}},
  \bibinfo{author}{\bibfnamefont{L.}~\bibnamefont{Balents}}, \bibnamefont{and}
  \bibinfo{author}{\bibfnamefont{M.~P.~A.} \bibnamefont{Fisher}},
  \bibinfo{journal}{Phys. Rev. B} \textbf{\bibinfo{volume}{58}},
  \bibinfo{pages}{1794} (\bibinfo{year}{1998}).

\bibitem[{\citenamefont{Yoshioka and Odintsov}(1999)}]{Yoshioka99}
\bibinfo{author}{\bibfnamefont{H.}~\bibnamefont{Yoshioka}} \bibnamefont{and}
  \bibinfo{author}{\bibfnamefont{A.~A.} \bibnamefont{Odintsov}},
  \bibinfo{journal}{Phys. Rev. Lett.} \textbf{\bibinfo{volume}{82}},
  \bibinfo{pages}{374} (\bibinfo{year}{1999}).

\bibitem[{\citenamefont{Nersesyan and Tsvelik}(2003)}]{Nersesyan03}
\bibinfo{author}{\bibfnamefont{A.~A.} \bibnamefont{Nersesyan}}
  \bibnamefont{and} \bibinfo{author}{\bibfnamefont{A.~M.}
  \bibnamefont{Tsvelik}}, \bibinfo{journal}{Phys. Rev. B}
  \textbf{\bibinfo{volume}{68}}, \bibinfo{pages}{235419}
  (\bibinfo{year}{2003}).

\bibitem[{\citenamefont{Balents and Fisher}(1997)}]{Balents97}
\bibinfo{author}{\bibfnamefont{L.}~\bibnamefont{Balents}} \bibnamefont{and}
  \bibinfo{author}{\bibfnamefont{M.~P.~A.} \bibnamefont{Fisher}},
  \bibinfo{journal}{Phys. Rev. B (R)} \textbf{\bibinfo{volume}{55}},
  \bibinfo{pages}{11973} (\bibinfo{year}{1997}).

\bibitem[{\citenamefont{Krotov et~al.}(1997)\citenamefont{Krotov, Lee, and
  Louie}}]{Krotov97}
\bibinfo{author}{\bibfnamefont{Y.~A.} \bibnamefont{Krotov}},
  \bibinfo{author}{\bibfnamefont{D.-H.} \bibnamefont{Lee}}, \bibnamefont{and}
  \bibinfo{author}{\bibfnamefont{S.~G.} \bibnamefont{Louie}},
  \bibinfo{journal}{Phys. Rev. Lett.} \textbf{\bibinfo{volume}{78}},
  \bibinfo{pages}{4245} (\bibinfo{year}{1997}).

\bibitem[{\citenamefont{Lin}(1998)}]{Lin98-1}
\bibinfo{author}{\bibfnamefont{H.-H.} \bibnamefont{Lin}},
  \bibinfo{journal}{Phys. Rev. B} \textbf{\bibinfo{volume}{58}},
  \bibinfo{pages}{4963} (\bibinfo{year}{1998}).

\bibitem[{\citenamefont{Bunder and Lin}(2008)}]{Bunder08}
\bibinfo{author}{\bibfnamefont{J.~E.} \bibnamefont{Bunder}} \bibnamefont{and}
  \bibinfo{author}{\bibfnamefont{H.-H.} \bibnamefont{Lin}},
  \bibinfo{journal}{Phys. Rev. B} \textbf{\bibinfo{volume}{78}},
  \bibinfo{pages}{035401} (\bibinfo{year}{2008}).

\bibitem[{\citenamefont{\textrm{L{\'o}pez Sancho}
  et~al.}(2001)\citenamefont{\textrm{L{\'o}pez Sancho}, Mu{\~n}oz, , and
  Chico}}]{Lopez01}
\bibinfo{author}{\bibfnamefont{M.~P.} \bibnamefont{\textrm{L{\'o}pez Sancho}}},
  \bibinfo{author}{\bibfnamefont{M.~C.} \bibnamefont{Mu{\~n}oz}}, ,
  \bibnamefont{and} \bibinfo{author}{\bibfnamefont{L.}~\bibnamefont{Chico}},
  \bibinfo{journal}{Phys. Rev. B} \textbf{\bibinfo{volume}{63}},
  \bibinfo{pages}{165419} (\bibinfo{year}{2001}).

\bibitem[{\citenamefont{Zhao and Mazumdar}(2004)}]{Zhao04}
\bibinfo{author}{\bibfnamefont{H.}~\bibnamefont{Zhao}} \bibnamefont{and}
  \bibinfo{author}{\bibfnamefont{S.}~\bibnamefont{Mazumdar}},
  \bibinfo{journal}{Phys. Rev. Lett.} \textbf{\bibinfo{volume}{93}},
  \bibinfo{pages}{157402} (\bibinfo{year}{2004}).

\bibitem[{\citenamefont{Alfonsi et~al.}(2010)\citenamefont{Alfonsi, Lanzani,
  and Meneghetti}}]{Alfonsi10}
\bibinfo{author}{\bibfnamefont{J.}~\bibnamefont{Alfonsi}},
  \bibinfo{author}{\bibfnamefont{G.}~\bibnamefont{Lanzani}}, \bibnamefont{and}
  \bibinfo{author}{\bibfnamefont{M.}~\bibnamefont{Meneghetti}},
  \bibinfo{journal}{New J. Phys.} \textbf{\bibinfo{volume}{12}},
  \bibinfo{pages}{083009} (\bibinfo{year}{2010}).

\bibitem[{\citenamefont{Yang et~al.}(2010)\citenamefont{Yang, Beavers, Wang,
  Jiang, Liu, Jin, Mercado, Olmstead, and Balch}}]{Yang10}
\bibinfo{author}{\bibfnamefont{H.}~\bibnamefont{Yang}},
  \bibinfo{author}{\bibfnamefont{C.~M.} \bibnamefont{Beavers}},
  \bibinfo{author}{\bibfnamefont{Z.}~\bibnamefont{Wang}},
  \bibinfo{author}{\bibfnamefont{A.}~\bibnamefont{Jiang}},
  \bibinfo{author}{\bibfnamefont{Z.}~\bibnamefont{Liu}},
  \bibinfo{author}{\bibfnamefont{H.}~\bibnamefont{Jin}},
  \bibinfo{author}{\bibfnamefont{B.}~\bibnamefont{Mercado}},
  \bibinfo{author}{\bibfnamefont{M.~M.} \bibnamefont{Olmstead}},
  \bibnamefont{and} \bibinfo{author}{\bibfnamefont{A.~L.} \bibnamefont{Balch}},
  \bibinfo{journal}{Angew. Chem. Int. Ed.} \textbf{\bibinfo{volume}{49}},
  \bibinfo{pages}{886} (\bibinfo{year}{2010}).

\bibitem[{\citenamefont{Konstantinidis}(2015{\natexlab{b}})}]{NPK15-1}
\bibinfo{author}{\bibfnamefont{N.~P.} \bibnamefont{Konstantinidis}},
  \bibinfo{journal}{Eur. Phys. J. B} \textbf{\bibinfo{volume}{88}},
  \bibinfo{pages}{167} (\bibinfo{year}{2015}{\natexlab{b}}).

\bibitem[{\citenamefont{Konstantinidis}(2016{\natexlab{b}})}]{NPK16}
\bibinfo{author}{\bibfnamefont{N.~P.} \bibnamefont{Konstantinidis}},
  \bibinfo{journal}{J. Phys.: Condens. Matter} \textbf{\bibinfo{volume}{28}},
  \bibinfo{pages}{016001} (\bibinfo{year}{2016}{\natexlab{b}}).

\bibitem[{\citenamefont{Machens et~al.}(2013)\citenamefont{Machens,
  Konstantinidis, Waldmann, Schneider, and Eggert}}]{Machens13}
\bibinfo{author}{\bibfnamefont{A.}~\bibnamefont{Machens}},
  \bibinfo{author}{\bibfnamefont{N.~P.} \bibnamefont{Konstantinidis}},
  \bibinfo{author}{\bibfnamefont{O.}~\bibnamefont{Waldmann}},
  \bibinfo{author}{\bibfnamefont{I.}~\bibnamefont{Schneider}},
  \bibnamefont{and} \bibinfo{author}{\bibfnamefont{S.}~\bibnamefont{Eggert}},
  \bibinfo{journal}{Phys. Rev. B} \textbf{\bibinfo{volume}{87}},
  \bibinfo{pages}{144409} (\bibinfo{year}{2013}).

\end{thebibliography}

\begin{appendix}

\section{Subblock with Six Non-Neighboring Pentagons}
\label{appendix:subblocksixnonneighboringpentagons}
The subblock of six pentagons shown in Fig. \ref{fig:isolatedpentagonssubcluster}(a) is the unit that causes frustration as part of a (5,5) armchair-type fullerene molecule. When the interpentagon coupling $J'=0$ the magnetization curve corresponds to that of an isolated pentagon, which has constant susceptibility up to saturation with all spins sharing the polar angle in an umbrella configuration (Fig. \ref{fig:isolatedpentagonssubcluster}(b)). For higher $J'$ and low fields nearest-neighbor spins that belong to different pentagons remain very close to antiparallel, while nearest-neighbor intrapentagon correlations start to deviate from their isolated pentagon value.
The central pentagon has the strongest exchange energy and the smallest net spin, as each of the other pentagons has two spins with only two nearest neighbors which are therefore freer to respond to the field.
Right above the low-field magnetization discontinuity all polar angles become less than $\frac{\pi}{2}$, therefore each spin has negative magnetic energy. This is a spin umbrella configuration directly evolving from the $J'=0$ configuration, however now the spins do not share a common polar angle.
After the second magnetization discontinuity the configuration is the most symmetric. The inner ring spins, which belong to the central pentagon, share a common polar angle, their nearest neighbors  that define another ring a different polar angle and so on for a total of four distinct polar angles, one for each ring. The pentagon in the middle has again the strongest exchange energy and the smallest net spin, while the other five pentagons have the same exchange energy and net spin.
The introduction of the interpentagon coupling therefore generates two magnetization discontinuities which survive up to $J'=J$.

\section{Zero Field Ground State Energy and Saturation Field for the (5,5) Armchair-Type Fullerene Molecules}
\label{appendix:zerofieldgroundstatesaturationfield}
\begin{figure}
\includegraphics[width=3.5in,height=2.5in]{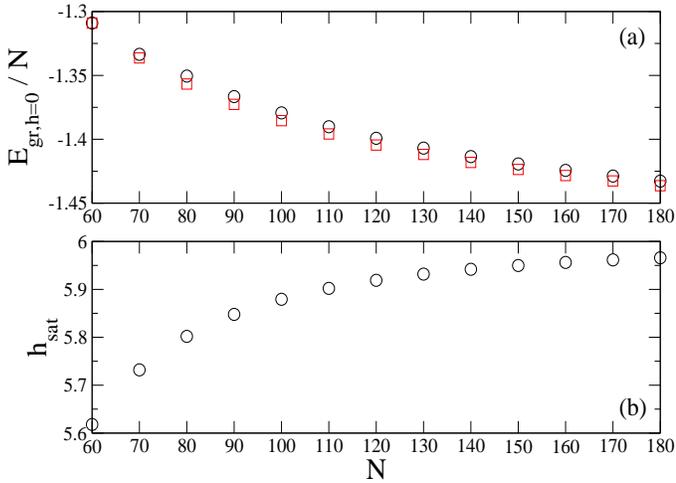}
\vspace{0pt}
\caption{(a) (Color online) The (black) circles show the zero field lowest energy per spin $\frac{E_{gr,h=0}}{N}$ as a function of the number of vertices $N$ for the (5,5) armchair-type fullerene molecules with $D_{5d}$ and $D_{5h}$ symmetry which are capped with six non-neighboring pentagons. The (red) squares show the sum of the energies of the isolated pentagons and hexagons per spin. $N=60$ corresponds to the $I_h$ symmetry molecule \cite{NPK07}. (b) Saturation field as a function of $N$.
}
\label{fig:D5dD5hzerofieldgroundstatesaturationfield}
\end{figure}

Nearest-neighbor spin directions in the zero-field ground state of an isolated pentagon have a relative angle of $\frac{4 \pi}{5}$. For an isolated hexagon nearest-neighbors are antiparallel. For the non-neighboring pentagon subblocks (Fig. \ref{fig:isolatedpentagonssubcluster}(a)) that cap the (5,5) armchair-type fullerene molecules with $D_{5d}$ and $D_{5h}$ symmetry no further frustration is introduced from the coupling of the pentagons, as it does not change the ground state relative angles of neighboring spins. If assembling pentagons and hexagons together to form a fullerene molecule with $N$ vertices would not introduce further frustration, the ground state energy should be equal to the sum of the ground state energies of the isolated pentagons and hexagons, which is $60cos\frac{4\pi}{5}-(\frac{3N}{2}-60)$. This is not the case for the (5,5) armchair-type fullerene molecules with $D_{5d}$ and $D_{5h}$ symmetry,
with their energy being slightly higher (Fig. \ref{fig:D5dD5hzerofieldgroundstatesaturationfield}(a)). On the other hand, the $N=60$ molecule which has $I_h$ symmetry and has the two caps directly connected achieves the lowest possible energy. The ground state energy per site $\frac{E_{gr,h=0}}{N}$ approaches asymptotically with $N$ the unfrustrated isolated hexagon value of $-\frac{3}{2}$. The only intrapentagon nearest-neighbor correlations equal to $cos\frac{4 \pi}{5}$ are the ones of the top and bottom pentagons (central pentagon in Fig. \ref{fig:isolatedpentagonssubcluster}(a)). In addition, the spins of the top and bottom pentagon are antiparallel to their nearest-neighbors, which belong to different pentagons. The saturation field for the molecules is given in Table \ref{table:D5dD5hsaturationfield} and plotted in Fig. \ref{fig:D5dD5hzerofieldgroundstatesaturationfield}(b), and approaches asymptotically with $N$ the value of 6.

\section{Ground State Spin Configuration Changes at the Discontinuities for the (5,5) Armchair-Type $D_{5d}$ Fullerene Molecules}
\label{appendix:groundstatechanges}
The magnetic field values for which the magnetization and susceptibility discontinuities of the (5,5) armchair-type $D_{5d}$ fullerene molecules occur, as well as the magnetization values below and above the magnetization discontinuities, are given in Tables \ref{table:D5dmagnetization} and \ref{table:D5dsusceptibility}.

\begin{figure}
\includegraphics[width=3.5in,height=2.5in]{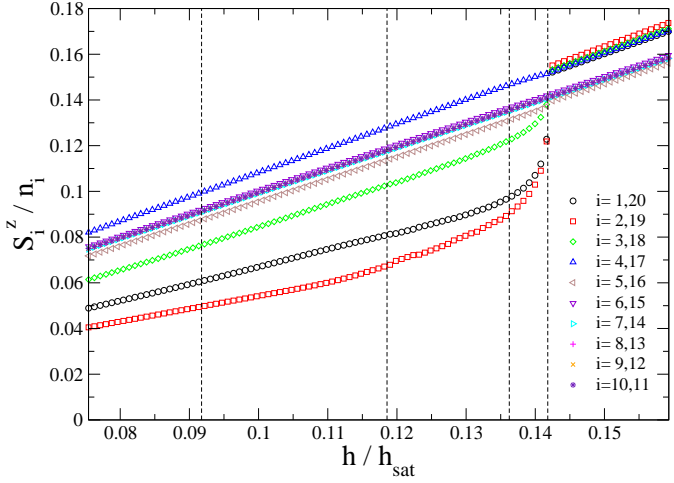}
\vspace{0pt}
\caption{(Color online) The different symbols (and colors) show the total magnetization per spin $\frac{S_i^z}{n_i}$ for each level of the spins according to their distance from one of the edges, given as a function of the magnetic field over its saturation value $\frac{h}{h_{sat}}$ for the (5,5) armchair-type fullerene molecule with $N=180$, which has $D_{5d}$ spatial symmetry. The number of spins per level $n_i$ equals 5 for $i=1,2,19,20$, and 10 for any other $i$. The dashed lines show the fields for the magnetization discontinuities. The values are symmetric with respect to the central plane of the molecule.
}
\label{fig:spinslevelszN=180lowfield}
\end{figure}

\begin{figure}
\includegraphics[width=3.5in,height=2.5in]{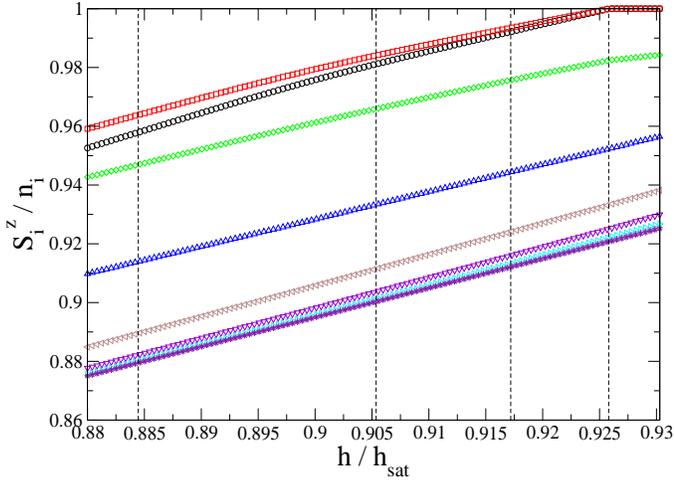}
\vspace{0pt}
\caption{(Color online) Total magnetization per spin $\frac{S_i^z}{n_i}$ for each level of the spins according to their distance from one of the edges, given as a function of the magnetic field over its saturation value $\frac{h}{h_{sat}}$ for the (5,5) armchair-type fullerene molecule with $N=180$, which has $D_{5d}$ spatial symmetry. The different symbols (and colors) follow Fig. \ref{fig:spinslevelszN=180lowfield}. The number of spins per level $n_i$ equals 5 for $i=1,2,19,20$, and 10 for any other $i$. The dashed line on the right shows the field for the susceptibility discontinuity, while the other three the fields for the magnetization discontinuities. The values are symmetric with respect to the central plane of the molecule.
}
\label{fig:spinslevelszN=180highfield}
\end{figure}

When the magnetic field is zero the spins of the top and bottom pentagon (central pentagon in Fig. \ref{fig:isolatedpentagonssubcluster}(a)) are antiparallel to their nearest-neighbors that belong to different pentagons (App. \ref{appendix:zerofieldgroundstatesaturationfield}). Once the field becomes non-zero two pairs are left that maintain their antiparallel directions, with one of the spins of the pair lined up with the field, and the other antiparallel with it. This shows that for low fields the spins belonging to the pentagons respond differently as part of the molecule, in comparison with the isolated six pentagon subblock (App. \ref{appendix:subblocksixnonneighboringpentagons}). When the field gets strong enough these pairs of spins cease to be antiparallel, and also parallel or antiparallel to the field. This change of their constant direction is accompanied by the low-field susceptibility discontinuity for $N \leq 140$, and by the lowest field magnetization discontinuity for $N=160$ and 180 (Fig. \ref{fig:spinslevelszN=180lowfield}). Similarly, in the ground state configuration just before saturation the spins at the top and bottom two levels of the molecules become aligned with the field (Fig. \ref{fig:spinslevelszN=180highfield}). This change to a constant direction occurs again with a discontinuity in the susceptibility.

The total magnetization $S_i^z$, $i=1,\dots,\frac{N}{10}+2$ for groups of spins that are at the same distance from the edges is the lowest for spins closer to the edges of the molecules for low fields (Fig. \ref{fig:spinslevelszN=180lowfield}). $i=1$ corresponds to the inner ring spins, which belong to the central pentagon (Fig. \ref{fig:isolatedpentagonssubcluster}(a)), $i=2$ to their nearest neighbors that define another ring and so on. In an open chain the individual spins respond to the field more strongly the closer to the edges they are \cite{Machens13}. Eventually the field makes the magnetization of these levels among the highest after the last of the low-field discontinuities, which is associated with the strongest change in magnetization in Fig. \ref{fig:D5dD5hmagnsusc}(e). Above this discontinuity all the spins have negative magnetic energy for the first time.

\section{Symmetry of the Ground State Spin Configurations for the (5,5) Armchair-Type $D_{5d}$ Fullerene Molecules}
\label{appendix:symmetrygroundstate}
The symmetry of the lowest classical energy configuration of the (5,5) armchair-type $D_{5d}$ fullerene molecules capped with two six-pentagon subblocks with isolated pentagons is determined from its nearest-neighbor correlations, with the requirement that a symmetry operation does not change their value between any two spins. The symmetry groups that leave the correlations unchanged are subgroups of $D_{5d}$ \cite{Altmann94} and are given in Table IV. The symmetry is highest for small fields and close to saturation. The zero field ground state spin configuration is unchanged for symmetry operations that belong to the $S_{10}$ group. When the field becomes non-zero the symmetry is reduced to $C_i$. Right above the second discontinuity the only symmetry operation is the identity. The symmetry group right above the highest-field discontinuity is $S_{10}$, while right below it it is $C_i$. The wider column includes the symmetry of the configuration that is the lowest energy state for the widest range of fields. It occurs in the middle of the field range, and it is the first ground state configuration where all spins have negative magnetic energy.

\end{appendix}


\begin{table}
\begin{center}
\caption{Saturation magnetic field for the (5,5) armchair-type fullerene molecules capped with six non-neighboring pentagons. The columns give the number of vertices and the saturation magnetic field. The $D_{5d}$ molecules have $N$ equal to an even multiple of 10, while the $D_{5h}$ molecules have $N$ equal to an odd multiple of 10. The $N=60$ molecule belongs to the $I_h$ symmetry group \cite{NPK07}. The data is plotted in Fig. \ref{fig:D5dD5hzerofieldgroundstatesaturationfield}(b).}
\begin{tabular}{c|c}
$N$ & $h_{sat}$ \\
\hline
 60 & $\frac{9+\sqrt{5}}{2}$ \\
\hline
 70 & 5.73205 \\
\hline
 80 & 5.80194 \\
\hline
 90 & 5.84776 \\
\hline
100 & 5.87939 \\
\hline
110 & 5.90211 \\
\hline
120 & 5.91899 \\
\hline
130 & 5.93185 \\
\hline
140 & 5.94188 \\
\hline
150 & 5.94986 \\
\hline
160 & 5.95630 \\
\hline
170 & 5.96157 \\
\hline
180 & 5.96595 \\
\end{tabular}
\label{table:D5dD5hsaturationfield}
\end{center}
\end{table}

\begin{table}
\begin{center}
\caption{Magnetization discontinuities for the (5,5) armchair-type $D_{5d}$ fullerene molecules capped on both sides with six pentagons not neighboring each other. The columns give the number of vertices, the magnetic field for which the discontinuity occurs with respect to the saturation field, the magnetization right below the discontinuity with respect to the saturated magnetization, and the corresponding value for the magnetization right above the discontinuity. The $N=60$ molecule belongs to the $I_h$ symmetry group \cite{NPK07}. The data is plotted in Figs. \ref{fig:D5dD5hmagnsusc}(a), \ref{fig:D5dD5hmagnsusc}(b) and \ref{fig:D5dD5hmagnsusc}(e).}
\begin{tabular}{c|c|c|c}
$N$ & $\frac{h}{h_{sat}}$ & $\frac{\hat{M}_{-}}{N}$ & $\frac{\hat{M}_{+}}{N}$ \\
\hline
 60 & 0.14692  & 0.11723    & 0.14790 \\
\hline
 60 & 0.94165  & 0.94574    & 0.94651 \\
\hline
 80 & 0.13391  & 0.1153196  & 0.1369667 \\
\hline
 80 & 0.85562  & 0.87382668 & 0.87397773 \\
\hline
 80 & 0.92235  & 0.9411153  & 0.9412365 \\
\hline
100 & 0.12544  & 0.1121451  & 0.1122307 \\
\hline
100 & 0.13624  & 0.123827   & 0.133446 \\
\hline
100 & 0.87742  & 0.89852222 & 0.89858053 \\
\hline
100 & 0.91318  & 0.9343398  & 0.9343756 \\
\hline
120 & 0.13508  & 0.125082   & 0.125743 \\
\hline
120 & 0.13958  & 0.13803    & 0.14309 \\
\hline
120 & 0.88966  & 0.91044946 & 0.91046701 \\
\hline
120 & 0.91169  & 0.9322856  & 0.9322932 \\
\hline
140 & 0.12906  & 0.1204307  & 0.1205064 \\
\hline
140 & 0.14067  & 0.136563   & 0.143995 \\
\hline
140 & 0.81364  & 0.83215130 & 0.83215559 \\
\hline
140 & 0.89669  & 0.9162739  & 0.9162788 \\
\hline
140 & 0.91305  & 0.93230792 & 0.93230936 \\
\hline
160 & 0.10540  & 0.0984430  & 0.0984548 \\
\hline
160 & 0.12282  & 0.1153954  & 0.1154144 \\
\hline
160 & 0.13977  & 0.134215   & 0.134328 \\
\hline
160 & 0.14146  & 0.139161   & 0.144565 \\
\hline
160 & 0.86579  & 0.8838119  & 0.8838135 \\
\hline
160 & 0.90128  & 0.9194573  & 0.9194584 \\
\hline
160 & 0.91522  & 0.9330304  & 0.9330306 \\
\hline
180 & 0.091760 & 0.08618793 & 0.08619039 \\
\hline
180 & 0.11857  & 0.1121242  & 0.1121302 \\
\hline
180 & 0.13624  & 0.1302571  & 0.1302697 \\
\hline
180 & 0.14180  & 0.14047    & 0.14469 \\
\hline
180 & 0.88446  & 0.90136320 & 0.90136369 \\
\hline
180 & 0.90536  & 0.92215730 & 0.92215752 \\
\hline
180 & 0.91719  & 0.93365306 & 0.93365312 \\
\end{tabular}
\label{table:D5dmagnetization}
\end{center}
\end{table}

\begin{table}
\begin{center}
\caption{Susceptibility discontinuities for the (5,5) armchair-type $D_{5d}$ fullerene molecules capped on both sides with six pentagons not neighboring each other. The columns give the number of vertices, and the magnetic field for which the discontinuity occurs with respect to the saturation field. The data is plotted in Figs. \ref{fig:D5dD5hmagnsusc}(a) and \ref{fig:D5dD5hmagnsusc}(b).}
\begin{tabular}{c|c}
$N$ & $\frac{h}{h_{sat}}$ \\
\hline
 80 & 0.1187 \\
\hline
 80 & 0.9546 \\
\hline
100 & 0.1179 \\
\hline
100 & 0.1364 \\
\hline
100 & 0.9400 \\
\hline
120 & 0.1121 \\
\hline
120 & 0.9333 \\
\hline
140 & 0.1078 \\
\hline
140 & 0.9296 \\
\hline
160 & 0.9273 \\
\hline
180 & 0.9258 \\
\end{tabular}
\label{table:D5dsusceptibility}
\end{center}
\end{table}

\begin{tabular}{c|c|c|c|c|c|c|c|c|c|c}
\multicolumn{11}{l}{\small TABLE IV: Symmetry of the lowest} \\
\multicolumn{11}{l}{\small energy spin configurations of the (5,5)} \\
\multicolumn{11}{l}{\small armchair-type fullerene molecules with} \\
\multicolumn{11}{l}{\small $D_{5d}$ spatial symmetry and capped with} \\
\multicolumn{11}{l}{\small six non-neighboring pentagons. The} \\
\multicolumn{11}{l}{\small first column is the number of vertices} \\
\multicolumn{11}{l}{\small $N$. The first symmetry column refers} \\
\multicolumn{11}{l}{\small to the case of zero field. The field} \\
\multicolumn{11}{l}{\small increases when going to the right.} \\
$N$ & \multicolumn{10}{c}{lowest energy} \\
    & \multicolumn{10}{c}{configuration symmetry} \\
\hline
 80 & $S_{10}$ & $C_i$ & $C_i$ & I & \multicolumn{4}{c|}{I} & $C_i$ & $S_{10}$ \\
\hline
100 & $S_{10}$ & $C_i$ & $C_i$ & I & I & \multicolumn{2}{c|}{I} & I & $C_i$ & $S_{10}$ \\
\hline
120 & $S_{10}$ & $C_i$ & I & I & \multicolumn{3}{c|}{I} & I & $C_i$ & $S_{10}$ \\
\hline
140 & $S_{10}$ & $C_i$ & I & I & \multicolumn{2}{c|}{$C_i$} & I & I & $C_i$ & $S_{10}$ \\
\hline
160 & $S_{10}$ & $C_i$ & I & I & $C_i$ & \textrm{ } $C_i$ \textrm{ } & I & I & $C_i$ & $S_{10}$ \\
\hline
180 & $S_{10}$ & $C_i$ & I & I & $C_i$ & \textrm{ } $C_i$ \textrm{ } & I & I & $C_i$ & $S_{10}$
\end{tabular}

\begin{figure}
\centerline{
\includegraphics[width=1.6in,height=1.4in]{Figure3}
\hspace{20pt}
\includegraphics[width=1.6in,height=1.4in]{Figure4}
}
\vspace{0pt}
\caption{Fullerene molecules that resemble (5,5) armchair nanotubes capped with a subblock of non-neighboring pentagons shown in Fig. \ref{fig:isolatedpentagonssubcluster}(a). (a) Molecule with $N=140$ vertices ($D_{5d}$ symmetry). It has a center of inversion and thicker bonds are located on the upper half, while thinner on the lower half of the molecule. (b) Molecule with $N=150$ vertices ($D_{5h}$ symmetry). Due to a symmetry plane the bonds of the lower half are directly below the bonds of the upper half of the molecule.
}
\label{fig:armchair}
\end{figure}

\begin{figure}
\centerline{
\includegraphics[width=1.6in,height=1.4in]{Figure9}
\hspace{20pt}
\includegraphics[width=1.6in,height=1.4in]{Figure10}
}
\vspace{0pt}
\caption{Fullerene molecules that resemble (5,0) zigzag nanotubes capped with a subblock of neighboring pentagons shown in Fig. \ref{fig:nonisolatedpentagonssubcluster}(a). (a) Molecule with $N=140$ vertices ($D_{5d}$ symmetry). It has a center of inversion and thicker bonds are located on the upper half, while thinner on the lower half of the molecule. (b) Molecule with $N=150$ vertices ($D_{5h}$ symmetry). Due to a symmetry plane the bonds of the lower half are directly below the bonds of the upper half of the molecule.
}
\label{fig:zigzag}
\end{figure}

\end{document}